\documentclass[sort&compress,preprint]{elsarticle}
\usepackage{geometry}
\usepackage{setspace}
\usepackage{amsmath,graphicx} 
\usepackage{siunitx}
\usepackage{braket}
\usepackage{amssymb}
\usepackage{gensymb}
\usepackage{caption}
\usepackage{subcaption}
\usepackage{mathtools}
\usepackage[thinc]{esdiff}
\usepackage[capitalise]{cleveref}
\usepackage{xfrac}
\usepackage[dvipsnames]{xcolor}
\usepackage[normalem]{ulem}
\usepackage{multirow}
\usepackage{leftidx}
\usepackage{mathrsfs}

\usepackage{float}
\usepackage{siunitx}

\journal{Journal of Nuclear Materials}

\begin{document}

\begin{frontmatter}
\onecolumn
\title{Impact of grain boundary and surface diffusion on predicted fission gas bubble behavior and release in UO$_2$ fuel}
\author[UF]{Md Ali Muntaha}
\author[UF]{Sourav Chatterjee}
\author[UTK]{Sophie Blondel}
\author[INL]{Larry Aagesen}
\author[LANL]{David Andersson}
\author[UTK,ORNL]{Brian D. Wirth}
\author[UF]{Michael R. Tonks\corref{mycorrespondingauthor}}
\cortext[mycorrespondingauthor]{Corresponding author}
\ead{michael.tonks@ufl.edu}

\address[UF]{Department of Materials Science and Engineering, University of Florida, Gainesville, FL 32611, USA}
\address[UTK]{Department of Nuclear Engineering, University of Tennessee, Knoxville, TN, 37996, USA}
\address[INL]{Computational Mechanics and Materials, Idaho National Laboratory, Idaho Falls, ID 83415, USA}
\address[LANL]{MST-8, Los Alamos National Laboratory, Los Alamos, NM, 87545, USA}
\address[ORNL]{Oak Ridge National Laboratory, Oak Ridge, TN, 37830, USA}

\begin{abstract}
In this work, we quantify the impact of grain boundary (GB) and surface diffusion on fission gas bubble evolution and fission gas release in UO$_2$ nuclear fuel using simulations with a hybrid phase field/cluster dynamics model. We begin with a comprehensive literature review of uranium vacancy and xenon atom diffusivity in UO$_2$ through the bulk, along GBs, and along surfaces. In our model we represent fast GB and surface diffusion using a heterogeneous diffusivity that is a function of the order parameters that represent bubbles and grains. We find that the GB diffusivity directly impacts the rate of gas release via GB transport, and that the GB diffusivity is likely below 10$^4$ times the lower value from Olander and van Uffelen (2001). We also find that the surface diffusivity impacts bubble coalescence and mobility, and that the bubble surface diffusivity is likely below $10^{-4}$ times the value from Zhou and Olander (1984). 
\end{abstract}


\end{frontmatter}
\newpage

\section{Introduction}
\label{sec:introduction}

In uranium dioxide (UO$_2$) fuel pellets for light water reactors, fission gases, primarily xenon (Xe) and krypton (Kr), are produced as byproducts of uranium nuclear fission. These fission gas atoms exhibit exceptionally low solubility within UO$_2$ grains \cite{Olander, Cacuci_book}, resulting in the formation of small intra-granular bubbles and larger intergranular bubbles at grain boundaries (GBs) and triple junctions (TJs) \cite{tonks,rest2019fission}. The intergranular bubbles at GBs and TJs are of particular concern, as their growth and interconnection create a percolated pathway for fission gases to be released from the fuel pellet to the fuel rod gap and plenum. There, the fission gas directly impacts fuel performance by increasing the plenum pressure and decreasing heat transport through the gap.

Fission gas release typically results from a three-stage process \cite{tonks}. In the first stage, gas atoms are generated within the bulk and are either trapped by intra-granular bubbles or diffuse to GBs. In the second stage, GB bubbles nucleate, grow, and coalesce until interconnected bubble networks allow gas to flow from GB bubbles to TJ tunnels. In the third stage, gas flows through interconnected TJ tunnels to release at free surfaces. Experimental evidence indicates that all three stages occur concurrently, with significant overlap between them \cite{Zacharie_98}. However, fission gas bubble evolution has never been directly observed during reactor operation so there are still many aspects of it that are not understood.

Mesoscale simulations of fission gas bubbles provide an alternative means of investigating their evolution, and the phase field method has emerged as the most popular approach for such simulations \cite{hu2009phase,Millet_2012_a,Larry_2019,PRUDIL2022153777}. One limitation of the phase field method is that it is too computationally expensive to resolve the small intragranular bubbles and GB bubble evolution in the same simulation. A recently developed model \cite{DongUk-Kim} overcame this limitation by coupling a phase field  model representing GBs and intergranular bubble evolution with spatially resolved cluster dynamics representing fission gas generation, diffusion, and intragraular bubble trapping. However, the original implementation of hybrid model did not consider fast fission gas diffusion along GBs and bubble surfaces, and neither has most other previous fission gas bubble phase field models \cite{hu2009phase, Larry_2019, Millet_2011, Millet_2012_a, Millet_2012_b, Millet_2012_c, Yulan-Li_2013, Xiao_2020}. In addition, it simulated an interior region of the fuel without a free surface, so it could not represent fission gas release.

In this work, we investigate the impact of fast diffusion along GBs and bubble surfaces on fission gas release using 2D and 3D simulations with a modified version of the hybrid model from Kim et al.~\cite{DongUk-Kim}. We modify it by adding fast GB and bubble surface diffusion and a free surface for fission gas release. The outline of this paper is as follows: first, we provide a detailed literature review of the vacancy and fission gas diffusivities values for UO$_2$ grains, GBs, and free surfaces; then,  we summarize the hybrid model and discuss the enhancements we have made; finally, we present and discuss our 2D and 3D simulation results.

\section{Survey of U vacancy and gas atom diffusivity values in UO$_2$}
\label{Diffusion Parameter Selection}
\begin{figure}
\centering
\begin{subfigure}{.48\textwidth}
  \centering
  \includegraphics[width=0.96\linewidth]{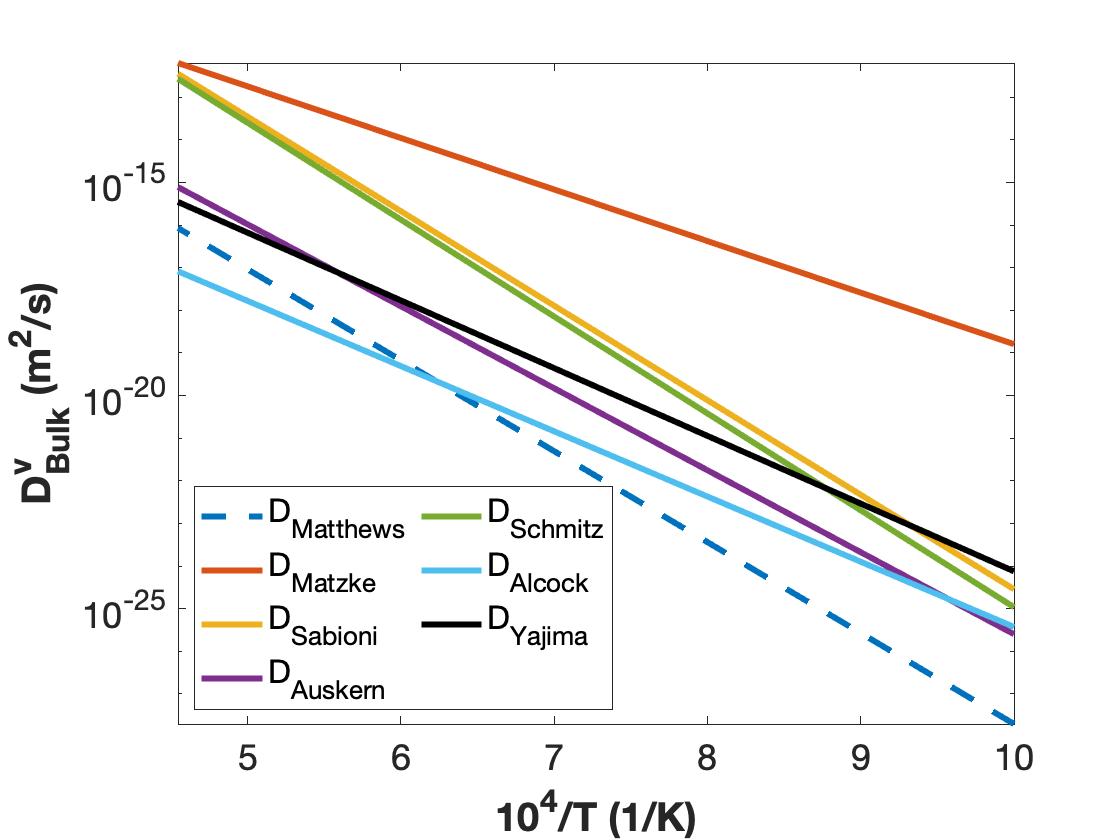}
  \caption{}
  \label{fig:U-vacancy diffusivity at bulk}
\end{subfigure}%
\begin{subfigure}{.48\textwidth}
  \centering
  \includegraphics[width=0.96\linewidth]{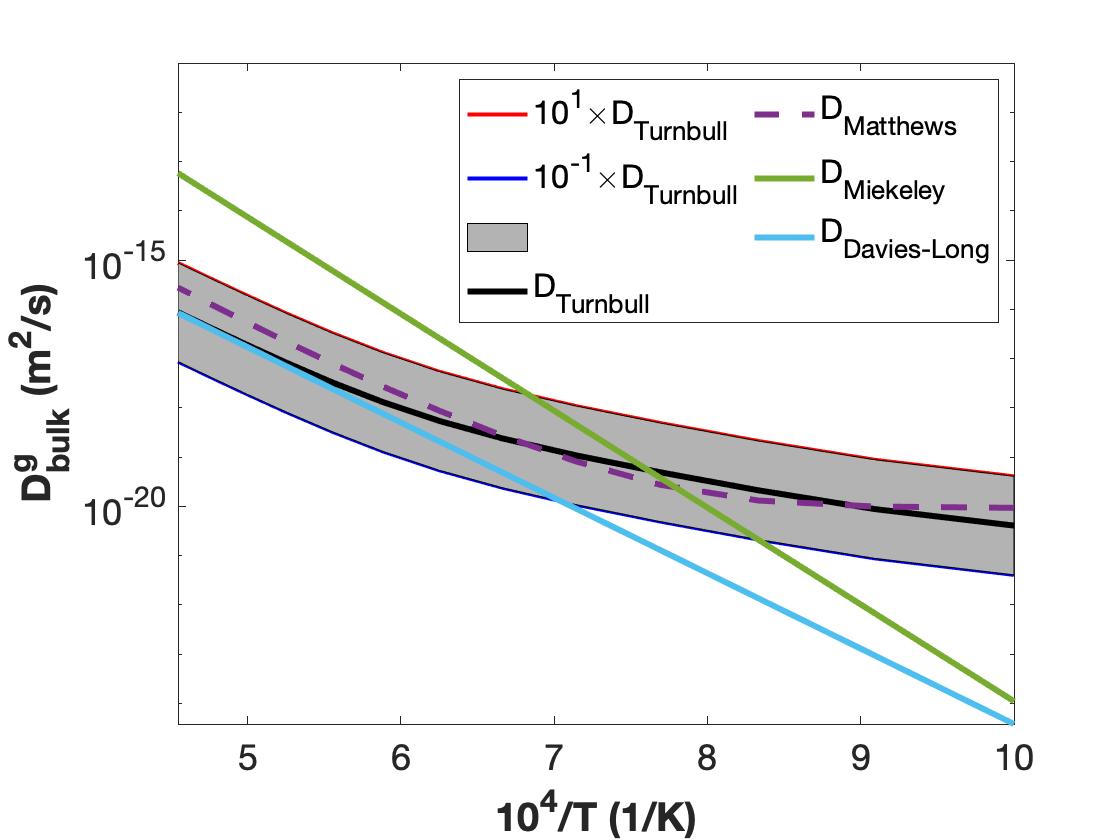}
  \caption{}
  \label{fig:Xe-gas diffusivity at bulk}
\end{subfigure} 
\begin{subfigure}{.48\textwidth}
  \centering

  \includegraphics[width=0.96\linewidth]{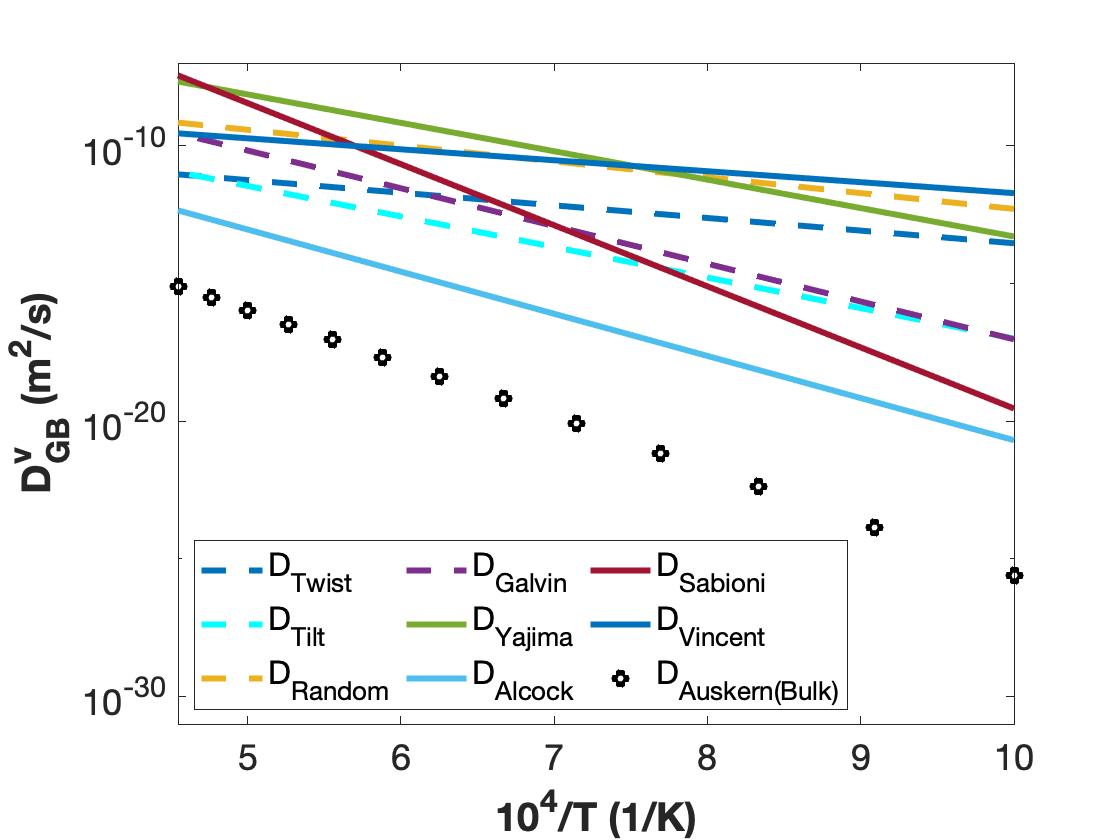}
  \caption{}
  \label{fig:U-vacancy diffusivity at GB}
\end{subfigure}
\begin{subfigure}{.48\textwidth}
  \centering
  \includegraphics[width=0.96\linewidth]{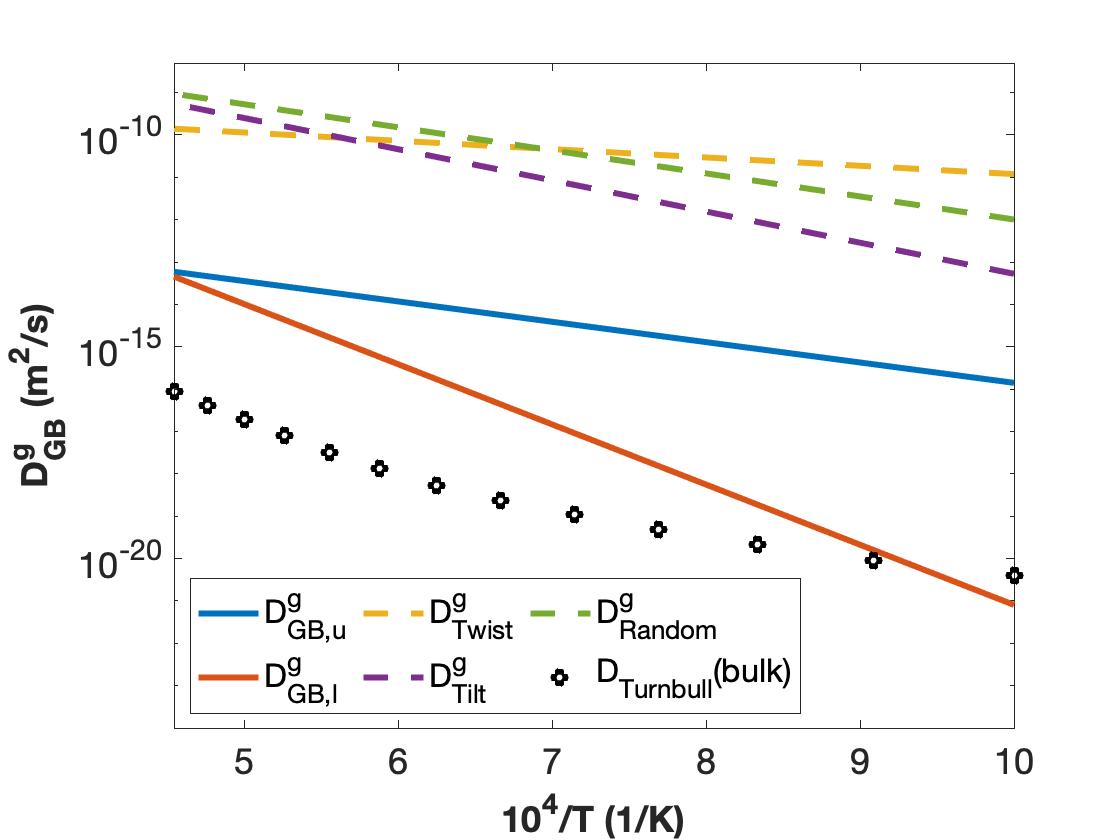}
  \caption{}
  \label{fig:Xe-gas diffusivity at GB}
\end{subfigure}
\begin{subfigure}{.48\textwidth}
  \centering
  \includegraphics[width=0.96\linewidth]{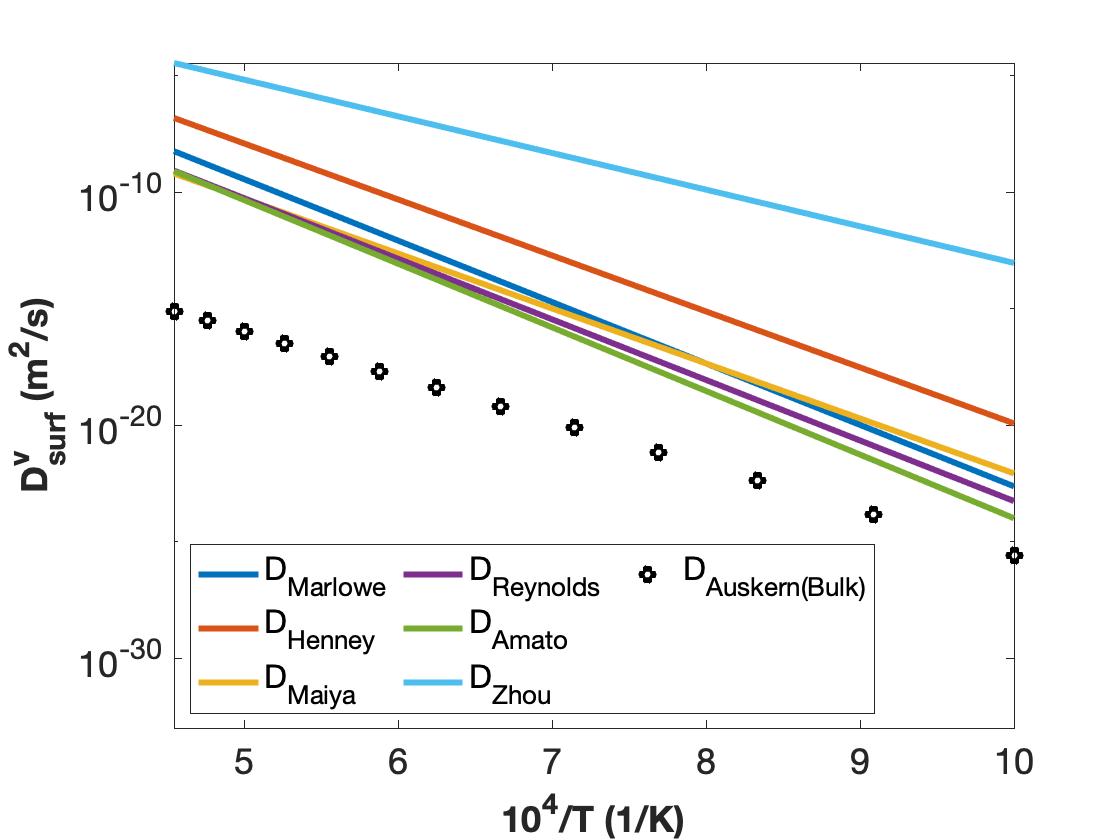}
  \caption{}
  \label{fig:U-vacancy diffusivity at surface}
\end{subfigure}%
\hspace{0.48\textwidth}
\caption{Diffusivity values for U vacancies and gas atoms in UO$_2$ from the literature. Bulk values through the crystal lattice are shown in (a) and (b), along GBs in (c) and (d), and vacancy diffusion along surfaces in (e). Experimental values are shown with solid lines, simulation values with dashed lines, and reference bulk diffusivities are shown with black diamonds.}
\label{fig:diffusivity-data-literature}
\end{figure}

The diffusivities of U vacancies and gas atoms have a direct impact on the fission gas release behavior in UO$_2$. Thus, their values have been measured in various experiments and calculated using simulation approaches. However, there is a large amount of scatter and uncertainty in these values. In Fig.~\ref{fig:diffusivity-data-literature}, we compare the bulk, GB, and surface diffusivities from the literature for temperatures ranging from 1000 K to 2200 K.

The values for the bulk diffusivity of U vacancies in UO$_2$, shown in Fig.~\ref{fig:U-vacancy diffusivity at bulk}, were obtained either by converting experimentally-measured values for U self diffusion by dividing by their reported vacancy concentration or by atomic-scale simulations. The measurements by Auskern and Belle~\cite{Auskern_1961}, Schmitz and Lindner~\cite{Schmitz_1966}, Alcock et al.~\cite{Alcock_1965} and Yajima et al.~\cite{Yajima_1966} were performed in the 1960s. Matzke~\cite{matzke_1987} measured the diffusivity in 1987, and obtained a value significantly higher than all other values. In 1998, Sabioni et al.~\cite{Sabioni_1998} used more modern approaches and obtained values very close to those from Schmitz and Lindner~\cite{Schmitz_1966}. Matthews et al.~\cite{Matthews_2019} used molecular dynamics simulations to calculate the vacancy diffusivity, resulting in an activation energy similar to that from Schmitz and Linder~\cite{Schmitz_1966} and Sabioni et al.~\cite{Sabioni_1998} but with a smaller prefactor. Overall, there is a large range of values; for a reference temperature $T = 1600$ K, the values range from $10^{-20}$ m$^2$/s to $10^{-13}$ m$^2$/s. If the value from Matzke is eliminated as an outlier, the values still vary by four orders of magnitude. This large variation is potentially due to the strong impact of non-stoichiometry and oxygen partial pressure on the U vacancy diffusivity \cite{ANDERSSON_2014_225} and to uncertainty in the reported vacancy concentrations used to convert from self diffusivities . Sabioni et al.~\cite{Sabioni_1998} observed a substantial effect of stoichiometry on uranium vacancy concentration. Their findings showed that the concentration for hypostoichiometric cases was around six orders of magnitude lower compared to the stoichiometric case. For our calculation of uranium vacancy diffusivity, we considered the stoichiometric concentration. The values from Auskern et al.~\cite{Auskern_1961} fall in the middle of the range if we neglect the value from Matzke, so we use it as a reference for the GB and surface diffusivity values.

The values for Xe diffusivity in single crystal UO$_2$, shown in Fig.~\ref{fig:Xe-gas diffusivity at bulk}, also come from experimental measurement and from simulations. Turnbull \cite{Turnbull} surveyed the available experimental data in 1982 and developed an overall fit that includes intrinsic and radiation enhanced diffusion. The scatter of the data, shown in gray in Fig.~\ref{fig:Xe-gas diffusivity at bulk}, is roughly plus or minus one order of magnitude and is much less than the scatter in the U vacancy data. The intrinsic diffusivity measured by Davies and Long \cite{Davies-long} was used in Turnbull's fit, which varies significantly from the value from Miekeley and Felix \cite{MIEKELEY1972297}. Matthews et al.~\cite{Matthews_2020} used a combination of density functional theory, molecular dynamics, and cluster dynamics to obtain Xe diffusivity values that also include intrinsic and radiation-enhanced diffusion and they compare well with the data surveyed by Turnbull. We use the Turnbull values as a reference for the Xe GB diffusivity.

In general, defect diffusion is faster along GBs than in the bulk, due to the additional free volume. Values for the GB diffusivity of U vacancies, shown in Fig.~\ref{fig:U-vacancy diffusivity at GB}, were again converted from U self diffusivities measured from experiments and taken from molecular dynamics simulations. The GB U self diffusivity values come from data obtained in the 1960s, from Alcock et al.~\cite{Alcock_1965} and Yajima et al.~\cite{Yajima_1966}, from 2000 by Sabioni et al.~\cite{Sabioni_2000}, and from 2009 by Vincent et al.~\cite{Vincent_2009}. Molecular dynamics simulations by Galvin et al.~\cite{Galvin_2023} were used to directly calculate the U vacancy GB diffusivity in UO$_2$. Liu et al.~\cite{Liu_2023} used the same approach as Galvin et al.\ to calculate diffusivities for twist, tilt, and random GBs. These calculated values fall within the range of the experimental values. The majority of these sources report GB diffusivity data ranging from $10^{-15}$ m$^2$/s to $10^{-8}$ m$^2$/s at $T = 1600$ K, which is approximately four to eight orders of magnitude greater than that of the bulk U vacancy diffusivity from Auskern et al.~\cite{Auskern_1961}.

Unfortunately, there are very few data on the Xe GB diffusivity in UO$_2$. It was measured by Olander and Van Uffelen \cite{Olander_GB_paper} and has been calculated by molecular dynamics simulations by Liu et al.~\cite{Liu_2023}, as shown in Fig.~\ref{fig:Xe-gas diffusivity at GB}. Olander and Van Uffelen \cite{Olander_GB_paper} used trace-irradiated UO$_2$ samples to measure Xe transport and estimate the GB diffusivity. However, due to the scatter in the bulk Xe diffusivity measurements, they estimated two GB diffusivities, one using the upper bulk diffusivity data \cite{MIEKELEY1972297}, and one using the lower bulk diffusivity data \cite{Davies-long}.  Liu et al. \cite{Liu_2023} computed GB data for twist, tilt, and random GBs using MD simulations. There are large differences between the measured and simulated values, though the migration energy from the simulations is similar to that from the upper measured values. The GB diffusivity ranges from $10^{-16}$ m$^2$/s to $10^{-9}$ m$^2$/s at $T = 1600$ K. When compared to bulk diffusivity data from Turnbull \cite{Turnbull}, xenon gas diffusivity at the GB was found to be three to eight orders of magnitude higher.

U vacancy diffusion is also accelerated at surfaces. Values for U Vacancy diffusivity along surfaces were again converted from U self diffusivity values, and are shown in Fig.~\ref{fig:U-vacancy diffusivity at surface}. Numerous researchers have reported U vacancy diffusivity data along surfaces.
Amato et al.~\cite{Amato_1966}, Reynolds \cite{Reynolds_1967}, Henney and Jones \cite{Henney_1968}, and Maiya \cite{maiya1971surface} employed mass transfer techniques (GB grooving and single scratch decay). Marlowe and Kaznoff \cite{Marlowe_1968} and Zhou and Olander \cite{Zhou_1984} relied on tracer techniques. However, tracers can migrate from the surface into the interior and it was suggested that this could be resulting in large underestimations of the surface diffusivities \cite{Robertson_1969,Olander_1981}. Ultimately, Zhou and Olander \cite{Zhou_1984} developed an approach that potentially avoids this issue. However, the value from Zhou and Olander is much higher than the values from mass transfer techniques. At $T=1600$ K, the mass transfer techniques give a value that is six orders of magnitude larger than the bulk U vacancy diffusivity from Auskern et al.~\cite{Auskern_1961} and the Zhou and Olander value is about twelve orders of magnitude larger. 

Due to the large variation in the surface and GB diffusivity values, it is important to understand their impact on the fission gas behavior. Simulations of the fission gas behavior that includes bulk, GB, and surface diffusion provide a useful means of studying these impacts. Therefore, we conduct a parametric study in Section \ref{sec:results} using the hybrid model developed by Kim et al.~\cite{DongUk-Kim} to examine the influence of GB and surface diffusivity on the microstructural evolution of fission gas bubbles and fission gas release in UO$_2$ nuclear fuel.

\section{Model formulation}
\label{sec:Model_formulation}
Our study is built upon the fission gas hybrid model \cite{DongUk-Kim} that predicts the evolution of both small intragranular bubbles and larger intergranular bubbles, as well as GB migration. This is accomplished by coupling a spatially-resolved cluster dynamics model, implemented in the Xolotl code \cite{blondel2017benchmarks}, with a phase field model, implemented using the MARMOT code \cite{Tonks_2012}. The cluster dynamics model in Xolotl represents the generation of Xe gas atoms due to fission reactions, their diffusion through the fuel matrix, gas atom clustering, and re-solution of gas atoms within the grains due to radiation and fission fragment damage. The phase field model in MARMOT includes GB evolution, intergranular bubble growth and coalescence, bubble migration, and gas atom diffusion along grain boundaries. At each time step, MARMOT passes updated GB and intergranular bubble surface positions to Xolotl and Xolotl passes the rate of gas arriving at these interfaces into MARMOT. The codes are coupled using capabilities from the MOOSE framework that MARMOT is based on \cite{gaston2015physics}. The computational cost of coupling the codes is small in 2D simulations and the hybrid model scales well on multiple processors. A more complete description of the hybrid model was provided by Kim et al.~\cite{DongUk-Kim}.

We have made several enhancements to the hybrid model to enable its use to investigate the impact of fast GB and bubble surface diffusion on fission gas release. The coupling between MARMOT and Xolotl is unchanged, but we made changes to the cluster dynamics model in Xolotl and the phase field model in MARMOT. In Xolotl, we modified the boundary conditions to represent gas release from grains at free surfaces and we modified the re-solution model. The cluster dynamics model is briefly summarized in Section~\ref{CD_model} and then we discuss these changes. In MARMOT, we changed the diffusivities of U vacancies and gas atoms along grain boundaries and at bubble surfaces to represent fast GB and surface diffusion, and we modified the boundary conditions to represent gas release from free surfaces. We summarize the phase field model and discuss the changes in Section~\ref{PF_Model}. 

\subsection{Cluster dynamics model summary and changes}
\label{CD_model}
For the hybrid fission gas model, the cluster dynamics model in Xolotl is used to model the transport and clustering of fission gas atoms within the UO$_2$ matrix. To simplify the model, it assumes that all fission gas atoms are Xe, the most common gaseous fission product. It simplifies the model by not directly representing vacancies but assuming vacancies are present with Xe atoms and clusters. The formation and evolution of intragranular bubbles is represented by modeling the evolution of the concentrations of Xe atom clusters using the expression
\begin{equation}
    \frac{\partial C_n}{\partial t} = D_n \nabla^2 C_n + \dot{F}y_n - Q(C_n),
\end{equation}
where $C_n$ is the concentration of a cluster containing $n$ Xe atoms and $D_n$ is the diffusion coefficient of the cluster. The Xe production is a function of the fission rate density $\dot{F}$ and the fission yield $y_n$ of the cluster per fission. $Q(C_n)$ describes the various reactions that impact the cluster concentration. Xolotl solves the coupled reaction-diffusion equations using the finite difference method with implicit time integration, where the PETSc library \cite{petsc-user-ref} is used to solve the nonlinear system of equations.

We assume that the single Xe atom is the only mobile cluster such that $D_{n>1}=0$. We also assume that fissions only yield single Xe atoms, such that $y_{n>1}=0$. Three types of reactions are considered according to 
\begin{equation}
    Q(C_n) = (k_n C_n C_1 - k_{n-1}C_{n-1}C_1) + (k^{emit}_n C_n - k^{emit}_{n+1} C_{n+1}) + (k^{reso}_n C_n - k^{reso}_{n+1} C_{n+1})
\end{equation}
where the first two terms represent absorption of a Xe atom to form a larger cluster $\mathrm{Xe}_1 + \mathrm{Xe}_n \rightarrow \mathrm{Xe}_{n+1}$, the second two terms represent Xe atom emission $\mathrm{Xe}_n \rightarrow \mathrm{Xe}_1 + \mathrm{Xe}_{n-1}$, and the last two terms represent Xe atom re-solution that is described by the same reaction equation as emission. The reaction rate expressions for absorption and emission are described by Kim et al.~\cite{DongUk-Kim}.

Previously \cite{DongUk-Kim}, re-solution was described using the heterogeneous resolution model \cite{setyawan2018atomistic} where a xenon bubble ejects an individual atom at a time. In this work, we use a different approach to model re-solution based on the Turnbull model described in \cite{pastore2023}: the xenon bubble is fully re-solved in the lattice, leading to a higher concentration of mobile xenon in the grains.

\subsection{Phase field model components and equations}
\label{PF_Model}

In the hybrid model, the phase field method in MARMOT is used to model GB migration, the interaction between GBs and gas bubbles, and intergranular bubble growth and coalescence. The phase field model formulation is similar to that described by Aagesen et al. \cite{Larry_2019} and it represents the evolving microstructure using variable fields. Grains are represented by order parameters denoted as $\eta_{1}, \eta_{2}, ..., \eta_{n}$. These order parameters have a value of 1 within corresponding grains, 0 in all other grains, and they smoothly transition from 0 to 1 across GBs. Similarly, an additional order parameter $\eta_0$ is designated to represent all the bubbles. We also represent Xe atoms and U vacancies using their respective chemical potentials: $\mu_{g}$ and $\mu_v$. We assume Xe atoms are substitutional on the U lattice, such that there is a U vacancy associated with each Xe atom. We neglect the vacancies and interstitials on the oxygen lattice and uranium interstitials since they migrate very quickly compared to gas atoms and U vacancies and are therefore not rate limiting. These variables evolve according to partial differential equations to minimize the overall grand potential of the system.

The model uses the grand potential formulation originally presented by Plapp \cite{Plapp_2011} and it is described in detail in Kim et al.~\cite{DongUk-Kim} and Aagesen et al.~\cite{Larry_2019}. Here, we provide a brief overview of the equations solved in the model. 
The order parameters describing the grains and bubbles are evolved according to the Allen-Cahn equation
\begin{equation}
    \frac{\partial{\eta_i}}{\partial{t}}=-L \frac{\delta{\Omega}}{\delta{\eta_i}},\ \mathrm{with}\ i = 0,1,..., n, \label{Eq:AC_evolution}
\end{equation}
where $L$ is the order parameter mobility, $\Omega$ is the grand potential, and $\frac{\delta}{\delta \eta_i}$ represents a variational derivative. $\Omega$ is written as function of the local grand potential density \cite{Larry_2018}
\begin{equation}
    \Omega =\int_{V}\left(m\left(\sum_{i=0}^{N}\left(\frac{\eta_i^4}{4}-\frac{\eta_i^2}{2}\right)+\sum_{i=0}^{N}\sum_{j \neq i}^{N}\frac{\gamma_{ij}}{2}\eta_{i}^2\eta_{j}^2+\frac{1}{4}\right)+\frac{\kappa}{2} \sum_{i=0}^{N} |\nabla\eta_{i}|^2+{ h_m \omega_m+h_b \omega_b}\right)dV, \label{eq:omega}
\end{equation}
where $m$ is the free energy barrier coefficient, $\gamma_{ij}$ is a model parameter that influences the shape of the diffuse interface and intergranular bubble contact angles, $\kappa$ is the gradient energy coefficient, $h_m$ and $h_b$ are switching functions that control matrix and bubble properties, respectively, and $\omega_m$ and $\omega_b$ are the grand potential densities of the matrix and bubble phases, respectively. The switching functions are defined as \cite{Moelans_2011}:
\begin{align}
    {h_m}&=\frac{\sum_{i=1}^{N}\eta_{i}^2}{{\sum_{i=0}^{N}\eta_{i}^2}}\\
    {h_b}&=1-{h_m}.
\end{align}
The grand potential densities of the matrix and bubble phase are given by \cite{Larry_2018, Plapp_2011}:
\begin{align}
    \omega_m &= f_{m,chem} - \mu_g \rho_g - \mu_v \rho_v, \label{omega_m_initial}\\
    \omega_b &= f_{b,chem} - \mu_g \rho_g - \mu_v \rho_v, \label{omega_b_initial}
\end{align}
where $f_{m,chem}$ and $f_{b,chem}$ are the chemical energy contributions of the Helmholtz free energies of each phase and $\rho_g$ and $\rho_v$ are the densities (defects per unit volume) of gas atoms and vacancies, respectively. 

For all GBs ($i,j>0$), $\gamma_{ij}=3/2$ to ensure the order parameters are symmetric across the interface \cite{Moelans_2008}. The values of $\gamma_{0j}$ and $\gamma_{i0}$ (at bubble surfaces) define the semi-dihedral angle that forms between a bubble surface and a GB. Some researchers have found that the semi-dihedral of intergranular bubbles in UO$_2$ is around $50^\circ$ \cite{White_2004}, while others report that it ranges from $40^\circ$ to $80^\circ$ \cite{hodkin1980ratio}. When $\gamma_{0j}=\gamma_{ij} = 3/2$, the semi-dihedral angle is $60^\circ$ degrees, which is within the experimental range, therefore we use a constant value $\gamma_{ij}=\gamma=3/2$ in the model.

The model parameters $L$, $m$, and $\kappa$ can be related to measurable material properties, the GB energy $\sigma_{GB}$ and the interface mobility $M_{int}$, and to the interfacial width $l_{int}$ to make the model quantitative \cite{Moelans_2011}: 
\begin{align}
    L_{GB}&={ \frac{4}{3} \frac{M_{int}}{l_{int}}}, \label{eq:L_GB}\\
    m&=\frac{6\sigma_{GB}}{l_{int}}\\
    \kappa &= \frac{3}{4} \sigma_{GB} l_{int}.
\end{align}
The interface mobility $M_{int}=M_{GB}$ for interfaces with $\eta_0=0$, where $M_{GB}=M_0 e^\frac{-Q}{R T}$ with the prefactor $M_0$ and activation energy Q. For bubble interfaces, where $\eta_0>0$, 
\begin{equation}
    M_{int} = M_{GB} (1 - g(\eta_0)) + M_{s} g(\eta_0),
\end{equation}
where $g(\eta_0) = \tanh(\frac{\eta_0}{0.2})$. The value of the surface mobility $M_s$ was set to be sufficiently high such that the bubble surface migration is diffusion controlled. 

The chemical free energy densities of the matrix phase $f_{m,chem}$ and the bubble phase $f_{b,chem}$ are approximated with parabolic functions \cite{Larry_2019} that are functions of the gas and vacancy concentrations ($c_v$ and $c_g$):
\begin{align}
f_{m,chem}&= \frac{1}{2} k_v^m\left(c_v-e^\frac{-E_v^f}{k_b T}\right)^2+ \frac{1}{2} k_g^m\left(c_g-e^\frac{-E_v^f}{k_b T}\right)^2, \label{eq:f_m_chem} \\
f_{b,chem}&= \frac{1}{2} k_v^b(c_v-c_v^{b,eq})^2+ \frac{1}{2} k_g^b(c_g-c_g^{b,eq})^2, \label{eq:f_b_chem}
\end{align}
where $k_v^i$ (for $i=m$ or $b$) is the parabolic coefficient for the phase with respect to vacancies and $k_g^i$ with respect to gas and $E_v^f$ and $E_g^f$ are the formation energies of a U vacancy and a Xe gas atom on a U lattice site, respectively. The initial equilibrium composition of vacancies and gas atoms in the bubble phase are $c_v^{b,eq}=0.546$ and $c_g^{b,eq}=0.454$ \cite{Larry_2019}. These initial equilibrium concentration results in the minimization of the parabolic free energy (Eq. \eqref{eq:f_b_chem}), which corresponds to the minimization of the Van der Waals free energy within the bubble phase. The chemical potentials values are related to the concentrations according to \cite{Larry_2019}:
\begin{align}
    \mu_g &= {V_a k_g^m}(c_g-c_g^{m,eq}), \label{eq:c_g_mu_g} \\
    \mu_v &= {V_a k_v^m}(c_v-c_v^{m,eq}), \label{eq:c_v_mu_v}
\end{align}
where $V_a = 0.0409$~nm$^3$ \cite{Idiri_2004} is the atomic volume of a U atom in the UO$_2$ crystal structure.

The evolution of the chemical potentials $\mu_g$ and $\mu_v$ is defined as \cite{Larry_2019}:
\begin{align}
 \frac{\partial{\mu_g}}{\partial{t}} &= \frac{1}{\chi_g} \left[\nabla.(D_g \chi_g \nabla\mu_g) + s_g - \sum_{i=0} ^{N} \frac{\partial\rho_g}{\partial\eta_{i}} \frac{{\partial\eta_{i}}}{\partial{t}} \right], \label{Eq:chem_g_evolution}\\
 \frac{\partial{\mu_v}}{\partial{t}} &= \frac{1}{\chi_v} \left[\nabla.(D_v \chi_v \nabla\mu_v) + s_v -  \sum_{i=0} ^{N} \frac{\partial\rho_v}{\partial\eta_{i}} \frac{{\partial\eta_{i}}}{\partial{t}} \right], \label{Eq:chem_v_evolution}
\end{align}
where $s_g$ and $s_v$ are the source terms for the production of Xe atoms and U site vacancies, $D_g$ and $D_v$ are the diffusion coefficient of the gas and vacancies, and $\chi_g$ and $\chi_v$ are the susceptibilities which are defined as \cite{Larry_2019},
\begin{align}
    \chi_g &= \frac{h_m}{V_a^2 k_g^m}+\frac{h_b}{V_a^2 k_g^b} \\
    \chi_v &= \frac{h_m}{V_a^2 k_v^m}+\frac{h_b}{V_a^2 k_v^b}.  
\end{align}

The gas source term $s_g$ is computed by Xolotl and passed to the phase field model. $s_g=0$ within a grain or bubble (when an order parameter equals 1), but $s_g>0$ at interfaces and its magnitude is equal to the total amount of gas diffusing to this location from the neighboring grid points, divided by the time step size. In the case of a moving interface, where the given location was situated in the grain at the previous time step and is now within the GB, the total amount of xenon (mobile and in bubbles) at this location is added to the previous quantity, divided by the time step. The latter quantity corresponds to the GB sweeping effect. Since Xolotl considers only gas atom clusters, it cannot define the vacancy source term $s_v$. Therefore, the vacancy source term is a function of the gas source term. As Xe diffuses through UO$_2$, it occupies a defect cluster containing multiple U-vacancies \cite{Matthews_2020}. At intermediate temperatures during reactor operation, Xe moves with four U-vacancies. Thus, it is assumed that $s_v$ = $4 s_g$ in the phase field model \cite{DongUk-Kim}.

\begin{figure}[tbp]
\centering
\includegraphics[width=0.4\linewidth]{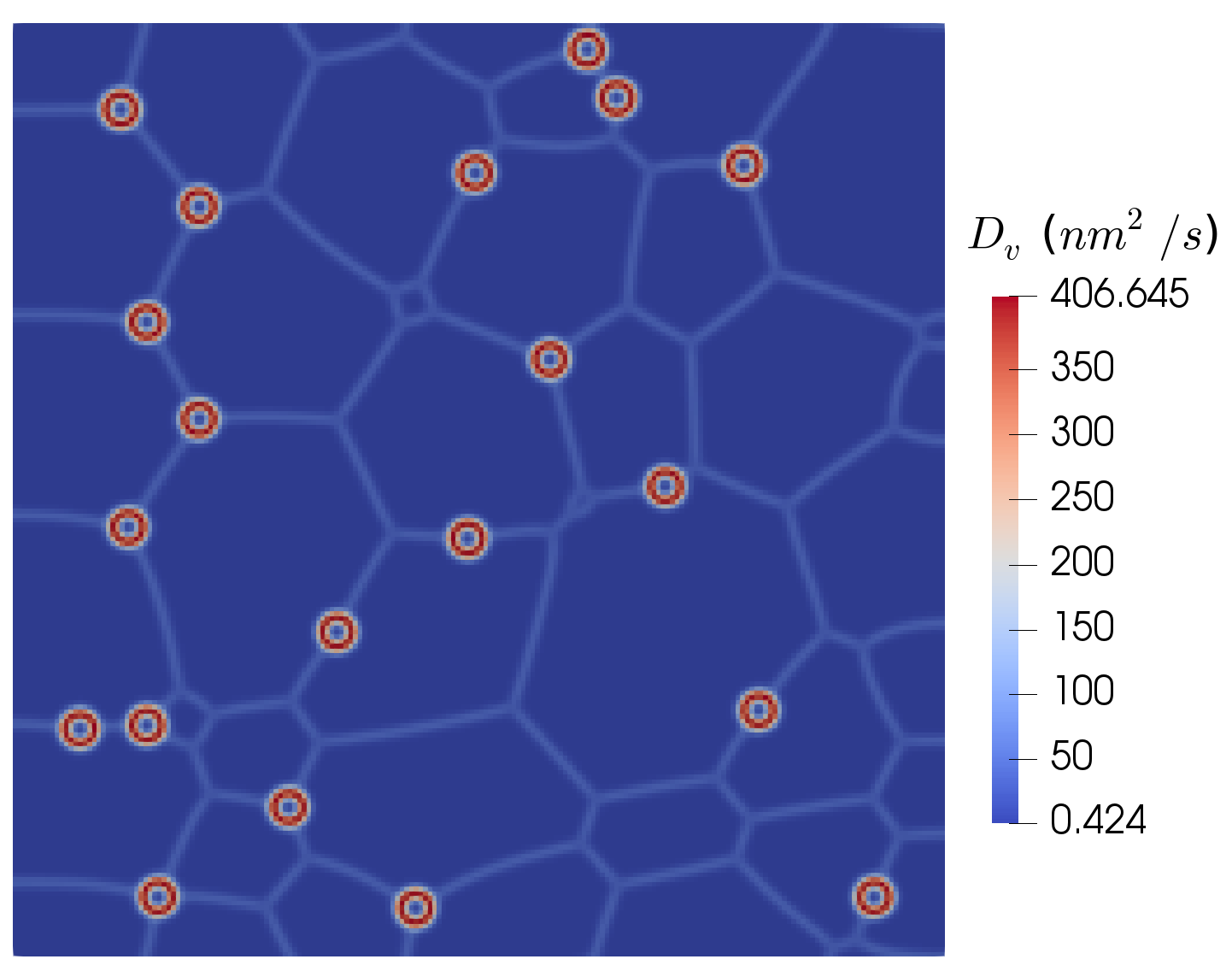}
\caption{Demonstration of the spatially varying diffusivity that accounts for fast GB and surface diffusion. The value of $D_v$ calculated using Eq.~\eqref{eq:uncorrected_D} is shown throughout a $30\ \mu m \times 30\ \mu m$ domain. Here, $D^v_{GB} = 100 D^v_{bulk}$ and $D^v_{surf} = 1000 D^v_{bulk}$}
\label{fig:test_case_demonstration_heterogeneity}
\end{figure}

In the fission gas phase field models described by Aagesen et al.~\cite{Larry_2019} and Kim et al.~\cite{DongUk-Kim}, the vacancy and gas diffusivities were assumed to be equal throughout the material. However, point defects diffuse faster along GBs and surfaces than within grains. Therefore, we modify the phase field model to make $D_v$ and $D_g$ functions of the order parameters $\eta_i$:
\begin{align}
    D_i&=\tilde{D}_{bulk}^i h_{bulk} + \tilde{D}_{GB}^i h_{GB} + \tilde{D}_{surf}^i h_{surf}, \label{eq:uncorrected_D}
\end{align} 
where $i=g$ or $v$ and $\tilde{D}_{bulk}^i$, $\tilde{D}_{GB}^i$, and $\tilde{D}_{surf}^i$ are the bulk, GB, and surface diffusivities used in the model, respectively. The interpolation functions that distinguish between bulk, surface, and GBs are defined as 
\begin{align}
    h_{gb} &= 16 \sum_{i=1}^{n} \sum_{j> i}^{n} \eta_{i}^2 \eta_{j}^2, \label{eq:h_gb} \\
    h_{surf} &= 16 \eta_0^2 (1-\eta_0)^2,\label{eq:h_{surface}} \\
    h_{bulk} &= 1 - h_{gb} - h_{surf}
\end{align} 
Figure \ref{fig:test_case_demonstration_heterogeneity} shows the value of $D_v$ from Eq.~\eqref{eq:uncorrected_D} across a 2D polycrystalline $30\ \mu m$ by $30\ \mu m$ domain with 20 fission gas bubbles and assuming $D^v_{GB} = 100 D^v_{bulk}$ and $D^v_{surf} = 1000 D^v_{bulk}$. Note the GB and surface diffusion for Xe atoms includes the Xe atom and the U vacancy where it resides. At a void surface, the Xe atom will not impact the diffusion rate significantly; thus, the vacancy will dominate the surface diffusion rate and we assume that U vacancy and Xe surface diffusivities are equal.

It is important to note that if measured values of the GB and surface diffusivities are used in Eq.~\eqref{eq:uncorrected_D}, the defect transport along the interfaces will be too large if the interface width $l_{int}$ is larger than the true width of such interfaces (on the order of 0.5 nm for GBs and 0.1 nm for oxide surfaces). Typically, phase field simulations use interface widths that are many of orders of magnitude larger than the true widths to reduce the computational cost of resolving the diffuse interface. Thus, the GB and surface diffusivities must be reduced. We derive an approach for reducing the GB and surface diffusivities for large interface widths based on the Hart approximation for the effective polycrystal diffusion coefficient $D_{eff}$ \cite{balluffi2005kinetics}: 
\begin{equation}
    D_{eff} = D_{bulk} + \frac{w_{int}}{G} (D_{int} - D_{bulk}), 
\end{equation}
where $w_{int}$ is the true interface width and $G$ is the average grain size. If we equate the Hart approximation for cases using the true interface width and a larger interface width $l_{int}$, we can solve for the reduced interface diffusivity
\begin{equation}
    \tilde{D}_{int} = D_{int} \frac{w_{int}}{l_{int}} + D_{bulk}  \left( 1- \frac{w_{int}}{l_{int}} \right).
    \label{eq:corrected_Dint}
\end{equation}
Note that a recent paper has presented alternative approaches for reducing the interface diffusivity \cite{simon2022mechanistic} that are more complex than the one we use here. 

The values used in our simulations for the various model parameters are shown in Table \ref{table: Parameters for phase field simulations}. The table also provides the reference for where the values were obtained. 
 \begin{table}[btp]
\renewcommand\arraystretch{2.0}
\caption{The parameters used for the phase field model. The reference that is the source of the value is provided, where applicable.}
\begin{center}
\begin{tabular}{m{6cm} m{6cm}  m{2cm}}
\hline\hline
Parameter  & Value  &  Reference \\ 
\hline
T                       &   1600K                               &                           \\
$M_{o}$                  &   2.14$\times$ $10^{-7}$ m$^4$/{Js}   & \cite{Tonks-PC-Jake_2021} \\
$Q$                     &   290 kJ/{mol}                        & \cite{Tonks-PC-Jake_2021} \\
$\gamma_{ij}$           &   1.5                                 & \cite{Larry_2019}         \\
$V_{a}$                 &   0.0409 nm$^3$                       & \cite{Idiri_2004}         \\
$\sigma_{GB}$           &   1.5 J/m$^2$                         & \cite{Tonks-PC-Jake_2021} \\
$l_{int}$               &   480 nm                              & \cite{DongUk-Kim}         \\
$w_{GB}$         &   0.5 nm                              & \cite{Yao_2017}         \\
$w_{s}$         &   0.1 nm                              & \cite{Hall_1987}         \\
$Y_{Xe}$                &   0.2156                              & \cite{International_Atomic_Energy_2017} \\
$E_v^f$                 &   3 eV                                & \cite{Yulan-Li_2013}      \\ 
$E_g^f$                 &   4.66 eV                             & \cite{Matthews_2020}      \\ 
$c_v^{b,eq}$            &   0.546                               & \cite{Larry_2019}         \\ 
$c_g^{b,eq}$            &   0.454                               & \cite{Larry_2019}         \\ 
$k_g^{m}=k_v^{m}$       &   4.81$\times$$10^{11}$ J/{m$^3$}     & \cite{Yulan-Li_2013}      \\ 
$k_g^{b}=k_v^{b}$       &   9$\times$$10^{11}$ J/{m$^3$}        & \cite{Yulan-Li_2013}      \\ 
$M_s$           &   5.616$\times$$10^{-18}$ m$^4$/{Js}   & \cite{Larry_2019}         \\ 
$S_v$                   &   4$\times$ $S_g$                     & \cite{Matthews_2020}         \\
\hline\hline
\end{tabular}
\end{center}

\label{table: Parameters for phase field simulations}
\end{table}

\section{Numerical methods}
\label{sec:numerical}
\begin{figure}[tbh]
\centering
\includegraphics[width=0.5\linewidth]{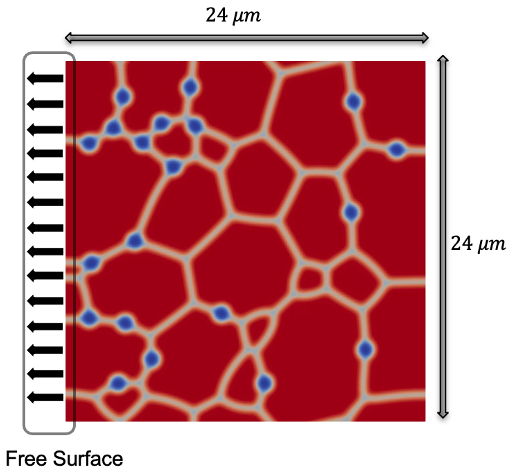}

\caption{The initial UO$_2$ microstructure used in our simulations. The microstructure has 25 initial grains, with a 5 $\mu$m initial average grain size, and 20 intergranular bubbles with a 480 nm initial radius. Periodic boundary conditions are applied on the top and bottom, the right boundary has a zero flux closed boundary condition, and the left boundary is considered a free surface.}
\label{fig:Domain_microstructure}
\end{figure}

In this work, we apply the hybrid fission gas model to investigate the impact of the surface and GB diffusivities on fission gas release using 2D polycrystal simulations. As described in Section \ref{sec:Model_formulation}, Xolotl uses the implicit finite difference method, MARMOT uses the implicit finite element method, and they are coupled using tools available in the MOOSE framework \cite{DongUk-Kim}.
In our investigation, we use a 24 $\mu$m by 24 $\mu$m polycrystal domain initially containing 25 grains with an average grain size of 5 $\mu$m. To avoid the need to model intergranular bubble nucleation, we begin the simulations with 20 circular bubbles with a 480 nm initial radius. The gas within the grains is initialized at zero. The initial grain and intergranular bubble structure is shown in Fig.~\ref{fig:Domain_microstructure}.

Both Xolotl and MARMOT use a uniform square grid with an element size/distance between points of 120 nm. First-order Lagrange elements are used in MARMOT. For the phase field model, a consistent interfacial width of $l_{int} = 480$ nm is used. Regarding boundary conditions, the top and bottom boundaries are treated as periodic, and a zero flux closed boundary condition is applied to the right boundary for all variables. The left boundary is considered an open or free surface. We represent the free surface in Xolotl using a Dirichlet boundary condition of zero for all cluster concentrations. We represent it in the phase field model using zero gradient boundary conditions for the grain order parameters and Dirichlet boundary conditions for the other variables with values of $\eta_0=0$, $\mu_v = 0$, and $\mu_g=-0.023$. The negative chemical potential for the gas atoms ensures that gas will only leave from the free surface and not be generated.

\begin{figure}[bth]
\centering
\includegraphics[width=0.6\linewidth]{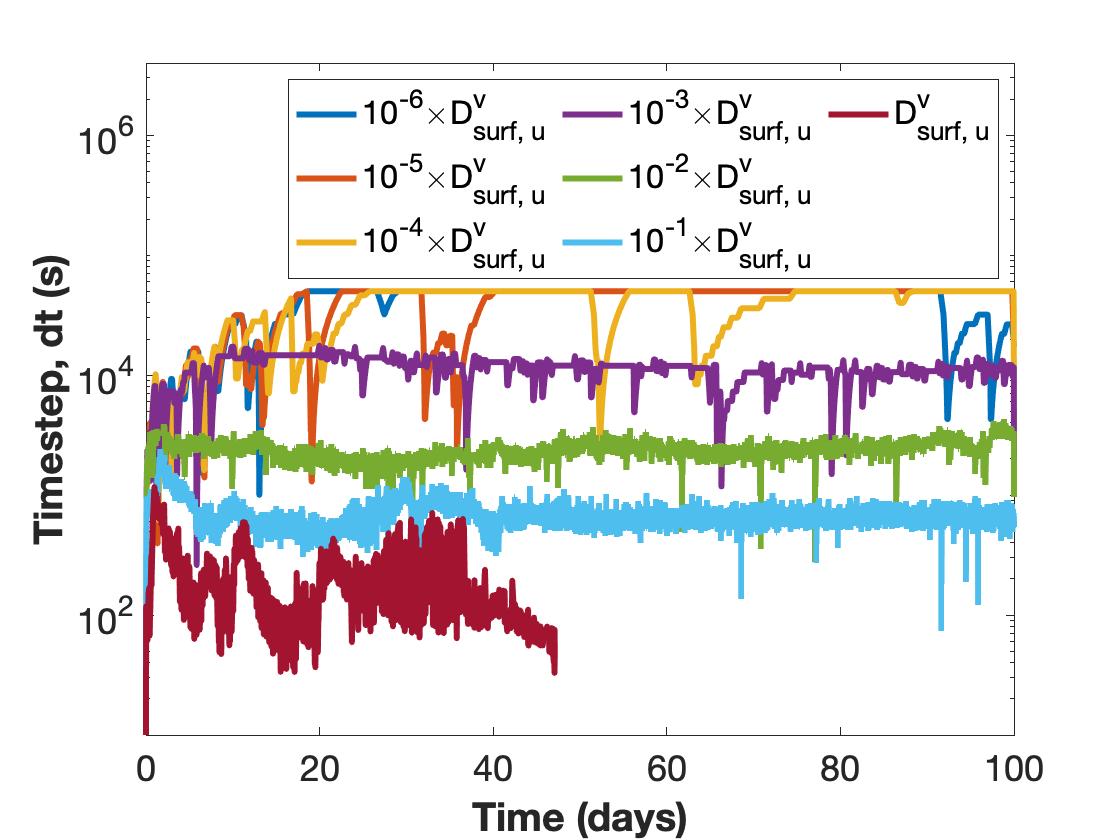} 
\caption{The impact of high interface diffusivity on the time step size. The surface diffusivity values are shown relative to the values from Zhou and Olander \cite{Zhou_1984}.}
\label{fig: dt}
\end{figure}

Adaptive time stepping is used in the simulations, where the initial time step is 10 s and the maximum time step used in the simulations is 50,000 s. The time step size is increased or decreased from one time step to the next to maintain an average of 20 nonlinear iterations in the MARMOT solve. Larger diffusivities result in faster evolution that requires a smaller time step to resolve. Therefore, simulations with larger GB and surface diffusivities require smaller time steps, and thus are more computationally expensive. This is demonstrated in Fig.~\ref{fig: dt} for a simulation at $T=1600$ K using the domain and initial conditions from Fig.~\ref{fig:Domain_microstructure}. The maximum time step used in the simulations decreases by many orders of magnitude as we increase the surface diffusivity, which has a large impact on the computational cost.  Due to this effect, we shorten the simulation time used for the fastest surface diffusivity investigated in our simulations to reduce the computational cost.

As shown in Figs.~\ref{fig:U-vacancy diffusivity at bulk} and \ref{fig:Xe-gas diffusivity at bulk}, the relationship between the bulk U vacancy and Xe diffusivity depends on the reference used for the U vacancy. The lower published values for U vacancy diffusivity are lower than the Xe diffusivity, while the Matzke \cite{matzke_1987} U vacancy diffusivity value is several orders of magnitude larger than the Xe diffusivity. In the models by Aagesen et al.~\cite{Larry_2019} and Kim et al.~\cite{DongUk-Kim}, it was assumed that bulk vacancy diffusivity was equal to the Xe atom diffusivity from Turnbull \cite{Turnbull}, which simplified the model and improved the convergence behavior. However, the impact of this assumption on the model predictions was not investigated. 

\begin{figure}[tbph]
\centering
\begin{subfigure}{.48\textwidth}
  \centering
  \includegraphics[width=0.96\linewidth]{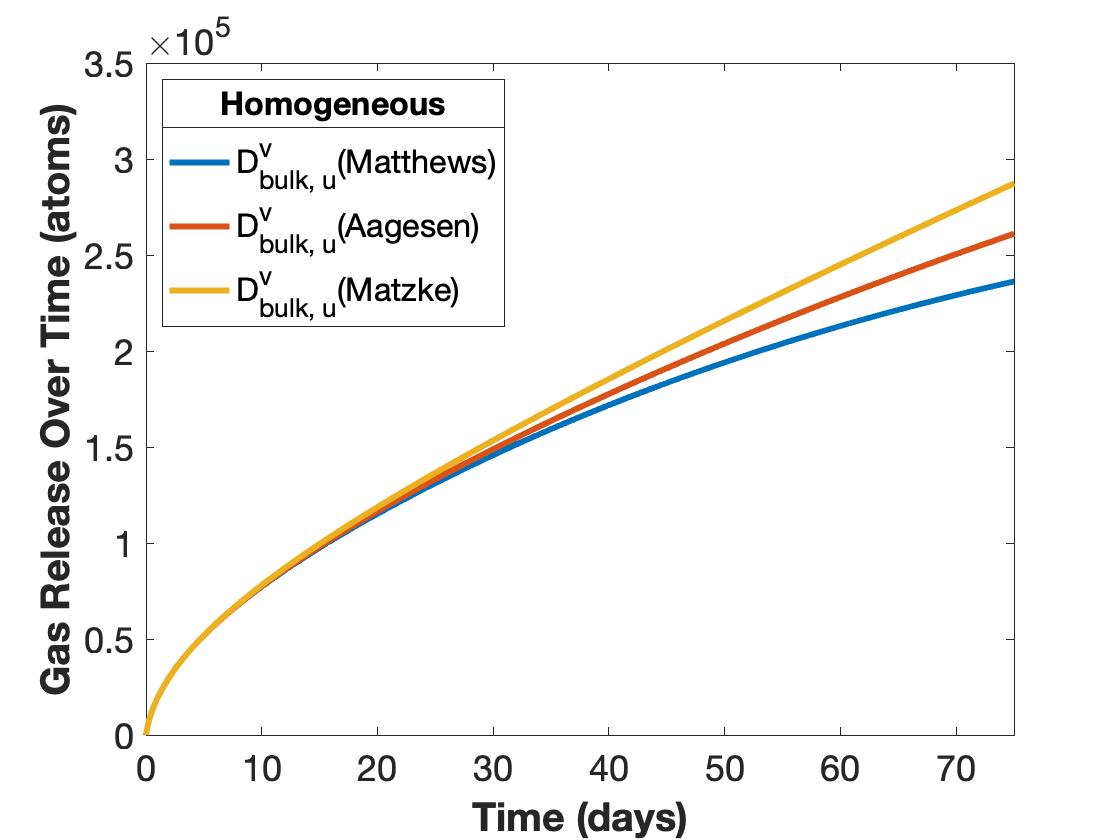}
  \caption{Homogeneous case}
  \label{fig:Gas Release Rate Over Time for Parametric bulk Homogeneous}
\end{subfigure}%
\begin{subfigure}{.48\textwidth}
  \vspace{0.5cm}
  \centering
  \includegraphics[width=0.96\linewidth]{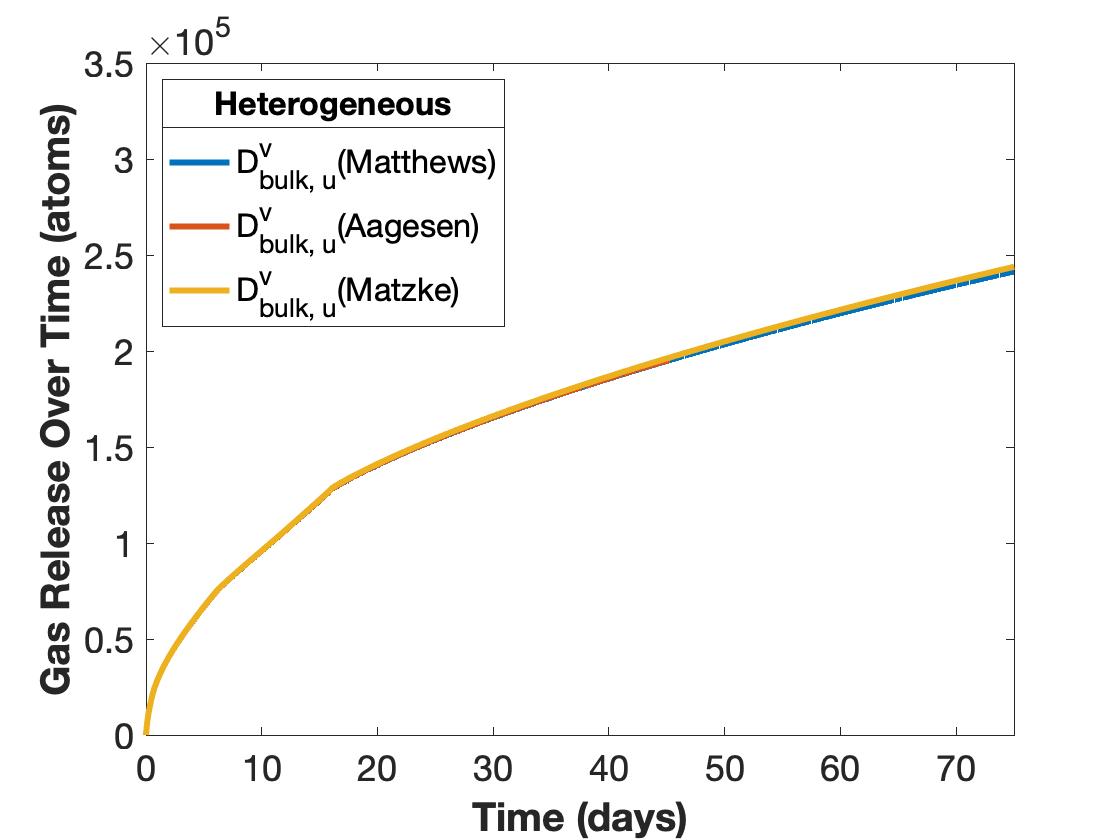}
  \caption{Heterogeneous case}
  \label{fig:Gas Release Rate Over Time for Parametric bulk Heterogeneous}
\end{subfigure}
\caption{The impact of U-vacancy bulk diffusivity on fission gas release, where we consider the values from Matthews et al.~\cite{Matthews_2019}, Matzke \cite{matzke_1987}, and assuming it is equal to the Xe diffusivity from Turnbull~\cite{Turnbull}, as done by Aagesen et al.~\cite{Larry_2019}. (a) shows the results for the homogeneous case when grain boundary and surface diffusion are equal to bulk diffusion. (b) shows the results when grain boundary and surface diffusivity are significantly greater than bulk diffusivity.}
\label{fig:Impact-of-Bulk-diffusivity}
\end{figure}

To quantify the impact of the bulk U vacancy diffusivity on the predicted fission gas release, we simulate the fission gas release at temperature $T=1600$ K and a fission rate $\dot{F} = 1.09 \times 10^{19}$ fissions/(m$^3$s) using the domain and initial condition shown in Fig.~\ref{fig:Domain_microstructure}. We find that when homogeneous diffusivity is considered, where vacancies and gas atoms have the same diffusivity throughout the bulk, GB, and surface, higher U vacancy diffusivity results in increased gas release over time even when the gas diffusivity is unchanged (Fig.~\ref{fig:Gas Release Rate Over Time for Parametric bulk Homogeneous}). However, when GB and surface diffusivity are several orders of magnitude higher than the bulk diffusivity (we assume $D_{GB}^v = 10^5 D_{bulk}^v$ and $D_{surf}^v = 10^7 D_{bulk}^v$), the impact of U vacancy diffusivity on fission gas release becomes negligible (Fig.~\ref{fig:Gas Release Rate Over Time for Parametric bulk Heterogeneous}). Therefore, for our investigation of the impact of fast GB and surface diffusion, we assume the bulk U vacancy diffusivity to be the same as the bulk Xe diffusivity, as was done by Aagesen et al.~\cite{Larry_2019} and Kim et al.~\cite{DongUk-Kim}. The impact of Xe gas bulk diffusivity on fission gas release was quantified and reported by Robbe et al.~\cite{Pieterjan_2021}. The fission gas release amount was reported to be identical when the Xe gas bulk diffusivity data was changed from that of Turnbull et al.~\cite{Turnbull} to Matthews et al.~\cite{Matthews_2020} for their temperature of study. Therefore, for the purposes of our parametric investigation, we have elected to use the Xe gas bulk diffusivity data from Turnbull et al.\cite{Turnbull}.

\section{Results}
\label{sec:results}
Due to the large uncertainty in the surface and GB diffusivity values from the literature, it is important to investigate their impact on the fission gas behavior. Thus, we carry out two parametric studies, one on the impact of the GB diffusivity on fission gas release and one on the impact of surface diffusivity. We use 2D simulations to reduce the overall computational expense. In both studies, we compare the gas release over time from the left boundary. The total gas release is computed by integrating the boundary gas flux over the side area (assuming a depth in the third dimension of 1 $\mu$m) and over time. We also compare the change in the number of bubbles, the number of grains, and the total computational wall time. In addition to these quantitative metrics, we also qualitatively compare the microstructure evolution over time.

\subsection{Impact of GB Diffusivity on Fission Gas Release}
\label{sec:Impact-of-GB-diffusivity}

To identify how GB diffusion impacts fission gas release, we carry out a parametric study in which we ignore fast surface diffusion and carry out simulations at increasing values of GB diffusivity. As shown in Figs.~\ref{fig:U-vacancy diffusivity at GB} and \ref{fig:Xe-gas diffusivity at GB}, at $T = 1600$ K the majority of literature-reported GB diffusivity data for both U vacancies and Xe lie between $10^{-15}$ m$^2$/s and $10^{-9}$ m$^2$/s, exhibiting a variation of up to six orders of magnitude. To simplify our parametric study, we assume that the GB vacancy and Xe diffusivities are equal. For the lowest GB diffusivity, we use the lower value from Olander and Van Uffelen \cite{Olander_GB_paper} (labeled $D^g_{GB,l}$). We then perform simulations with progressively higher GB diffusivities, increasing by orders of magnitude up to six orders of magnitude higher than $D^g_{GB,l}$. Thus, our simulations roughly encompass the full range of literature-reported GB diffusivity.
\begin{figure}[tbp]
\centering
\begin{subfigure}{.49\textwidth}
  \centering
  \includegraphics[width=0.99\linewidth]{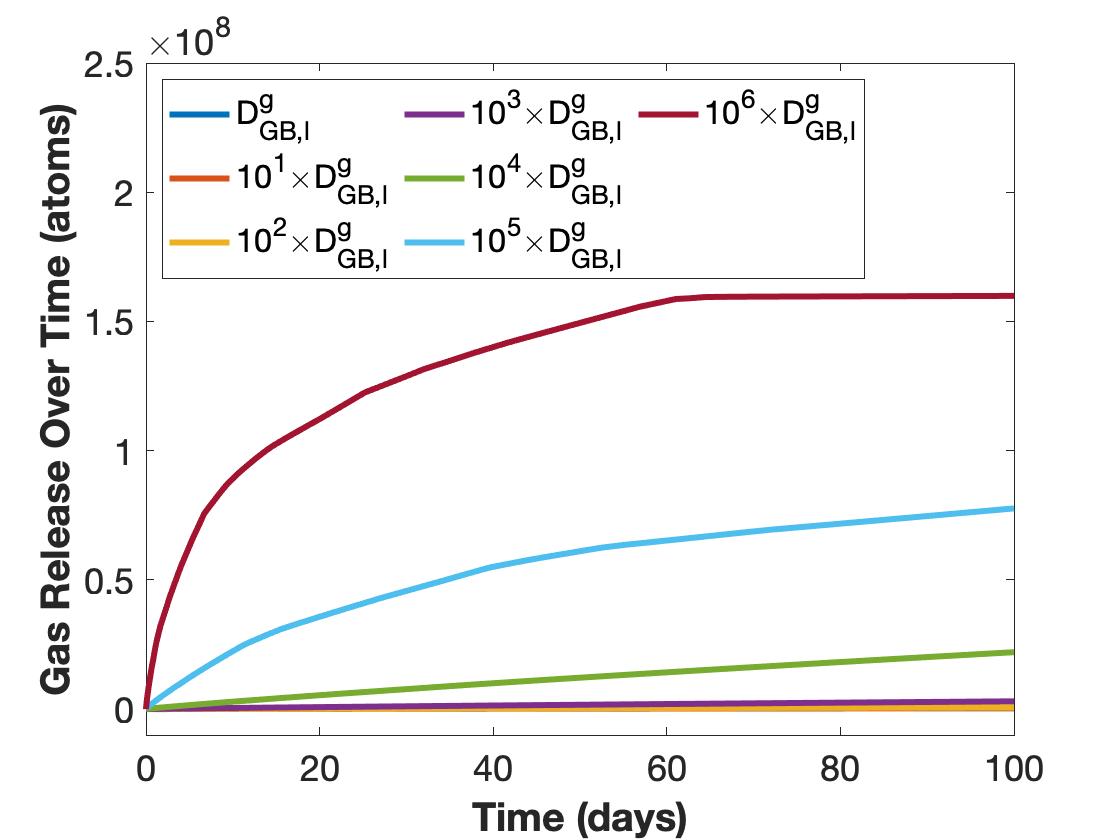}
  \caption{}
  \label{fig:Impact_GB_Gas_Release_Rate_Over_Time}
\end{subfigure}%
\begin{subfigure}{.49\textwidth}
  \vspace{0.5cm}
  \centering
  \includegraphics[width=0.99\linewidth]{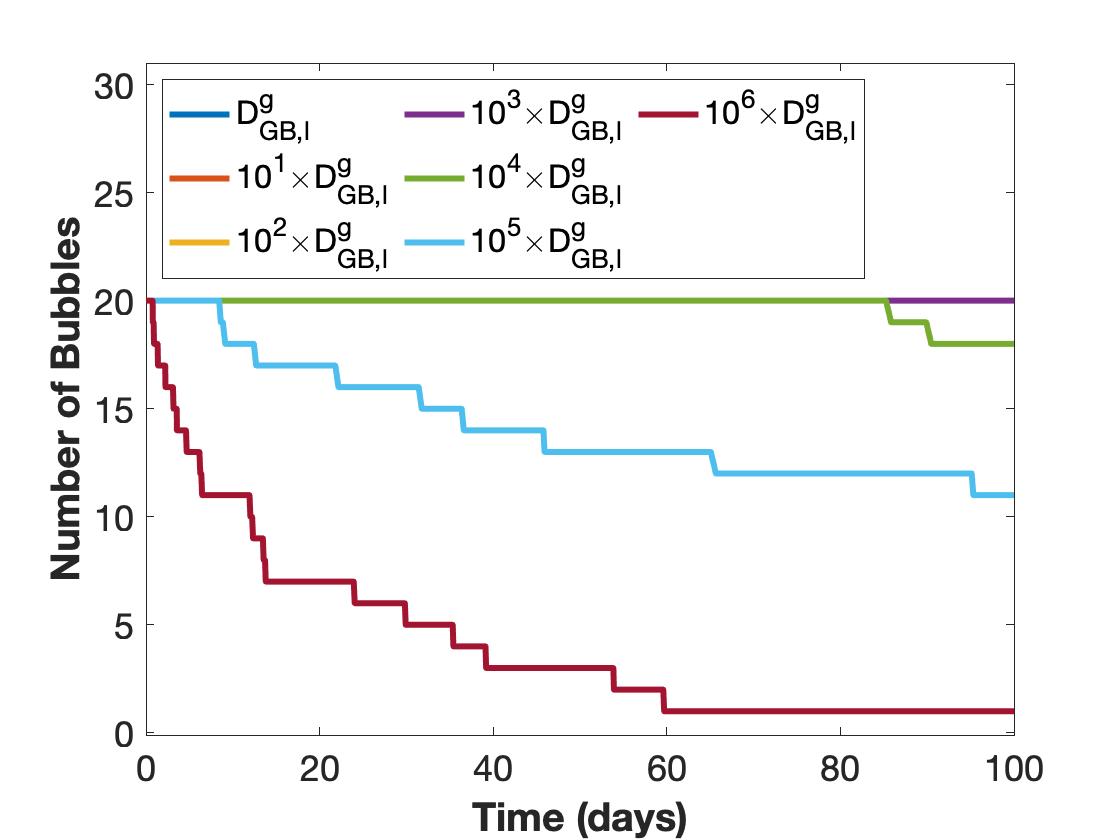}
  \caption{}
  \label{fig:Impact_GB_Number_of_bubbles}
\end{subfigure}%

\begin{subfigure}{.49\textwidth}
  \vspace{0.5cm}
  \centering
  \includegraphics[width=0.99\linewidth]{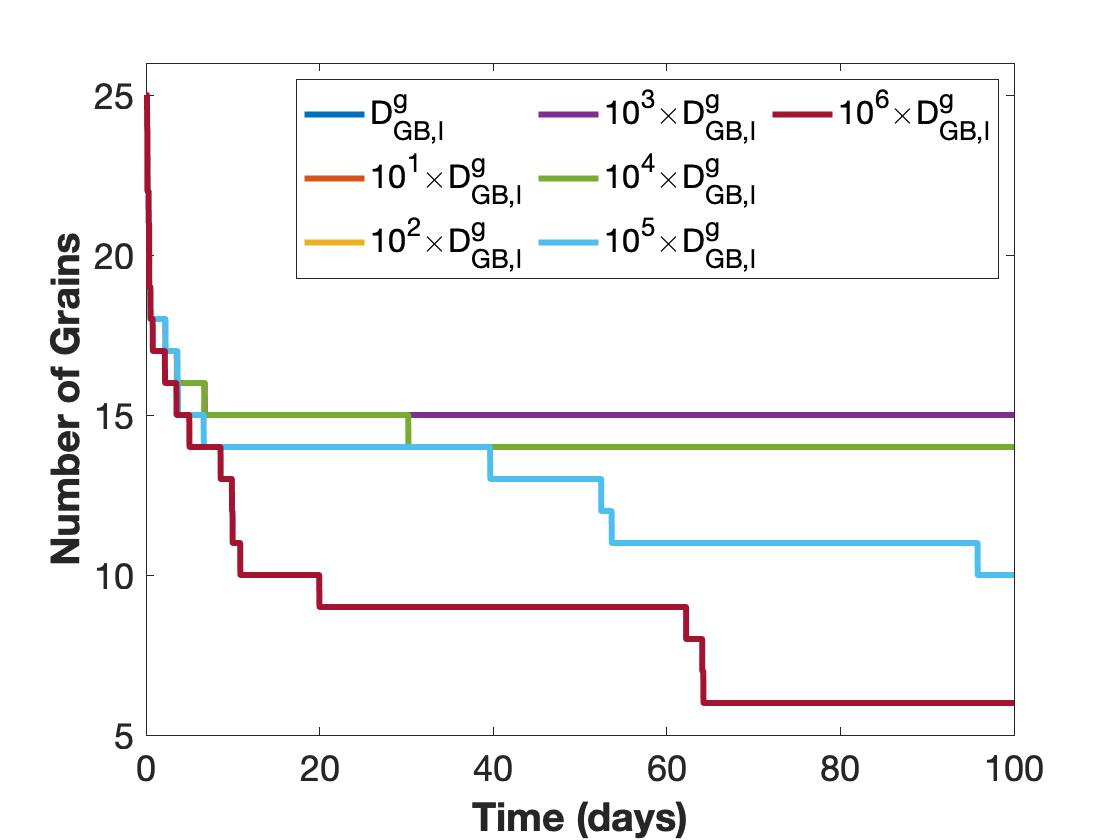}
    \caption{}
  \label{fig:Impact_GB_Number_of_grains}
\end{subfigure}
\begin{subfigure}{.49\textwidth}
  \vspace{0.5cm}
  \centering
  \includegraphics[width=0.99\linewidth]{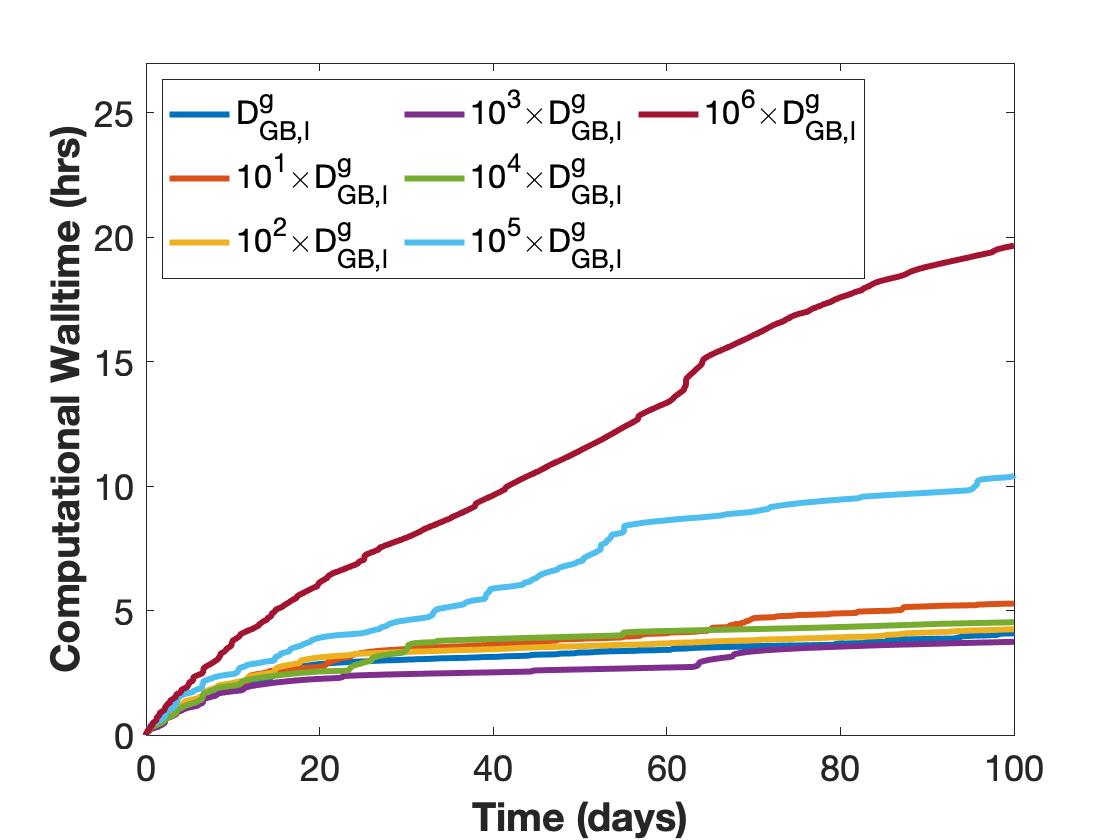}
  \caption{}
  \label{fig:Impact_GB_Computational_wall_time}
\end{subfigure}
\caption{The effects of GB diffusivity on fission gas release, where (a) shows the rate of gas release over time, (b) the number of bubbles, (c) the number of grains, and (d) the computational wall time. A homogeneous case with no fast GB diffusivity is shown for reference.}
\label{fig:Impact-of-GB-diffusivity}
\end{figure}

As shown in \cref{fig:Impact_GB_Gas_Release_Rate_Over_Time}, the GB diffusivity has a large impact on the gas release. As GB diffusivity increases, more gas travels along the GBs and escapes from the free surface. The gas release becomes significant enough to cause bubble collapse, reducing the number of bubbles (see \cref{fig:Impact_GB_Number_of_bubbles}), for GB diffusivities $\geq 10^4 D_{GB,l}$. The number of grains decreases rapidly at the start of the simulations due to GB migration, as shown in \cref{fig:Impact_GB_Number_of_grains}, until they are pinned by fission gas bubbles. For GB diffusivities $\leq 10^3 D_{GB,l}$, the number of grains reaches 15 and does not decrease further, since the bubbles do not evolve. For GB diffusivities $\geq 10^4 D_{GB,l}$, bubble evolution allows for additional GB migration. For GB diffusivities $\geq 10^5 D_{GB,l}$, bubble collapse results in significant reduction in the number of grains. Yet, as the number of grains goes down, the fission gas release slows. The faster diffusion and additional GB migration requires a smaller time step, which results in larger wall time, as shown in \cref{fig:Impact_GB_Computational_wall_time}. The total wall time (computational cost) is nearly four times greater with a GB diffusivity of $10^6 D_{GB,l}$ than for $D_{GB,l}$.

\begin{figure}[!htbp]
\centering
\begin{subfigure}{\textwidth}
  \centering
  \includegraphics[width=0.99\linewidth]{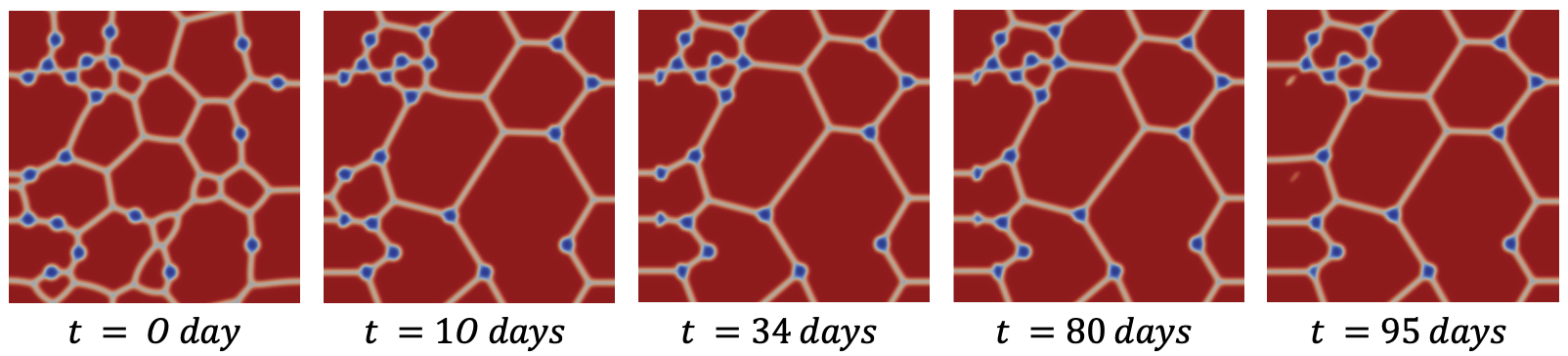}
  \caption{$D_{GB}=10^4\times D^g_{GB,l}$}
  \label{fig:Parametric GB Microstructure DG_1e4_D_GB_l}
\end{subfigure}%

\begin{subfigure}{\textwidth}
  \vspace{0.25cm}
  \centering
  \includegraphics[width=0.99\linewidth]{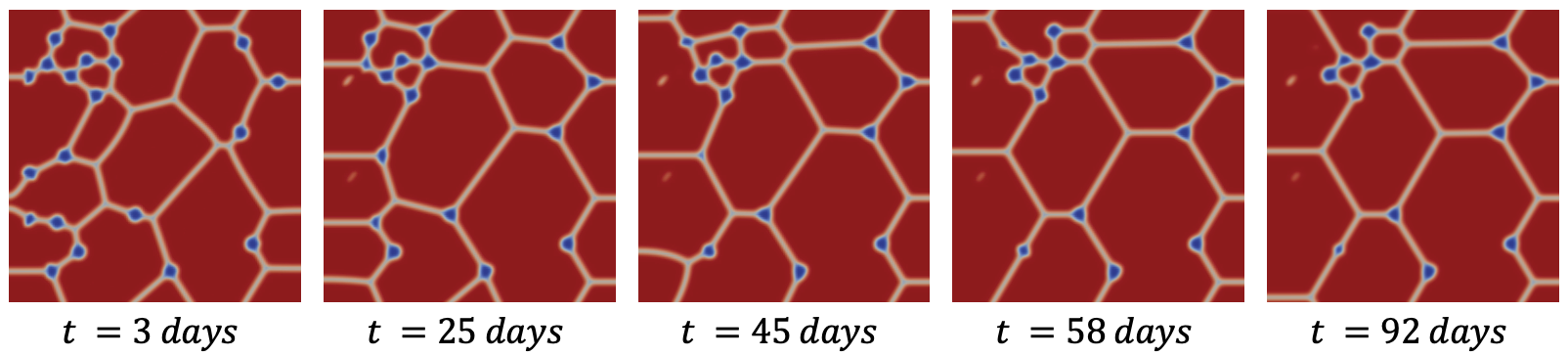}
  \caption{$D_{GB}=10^5\times D^g_{GB,l}$}
  \label{fig:Parametric GB Microstructure DG_1e5_D_GB_l}
\end{subfigure}%

\begin{subfigure}{\textwidth}
  \vspace{0.25cm}
  \centering
  \includegraphics[width=0.99\linewidth]{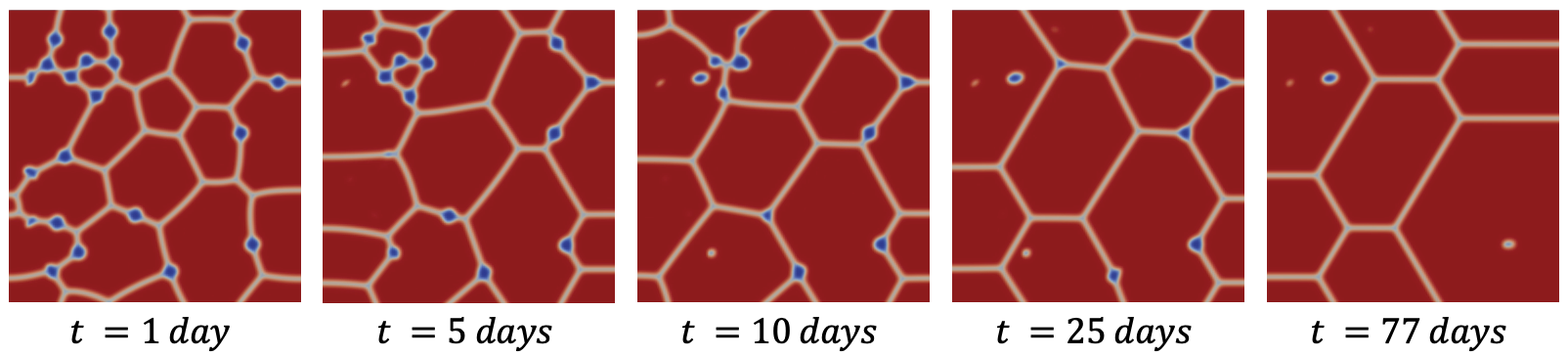}
  \caption{$D_{GB}=10^6\times D^g_{GB,l}$}
  \label{fig:Parametric GB Microstructure DG_1e6_D_GB_l}
\end{subfigure}

\caption{The impact of GB diffusivity on microstructure evolution. Snapshots of the microstructure are shown at different times for grain boundary diffusivity: (a) $10^4\times D^g_{GB,l}$, (b) $10^4\times D^g_{GB,l}$, and (c) $10^4\times D^g_{GB,l}$. The times are selected for each snapshot to best illustrate the microstructure change with the different GB diffusivities. The simulation's initial microstructure is shown in Fig.~\ref{fig:Domain_microstructure}. Grains are shown in red, GBs in yellow, and bubbles in blue.}
\label{fig:Impact-of-GB-diffusivity-microstructure}
\end{figure}

The relationship between bubble collapse, grain growth, and fission gas release is more clear from a qualitative analysis of the microstructure evolution, shown in \cref{fig:Impact-of-GB-diffusivity-microstructure}. For a GB diffusivity of $10^4 D_{GB,l}$, the two bubbles that disappear are the closest to the free surface and are connected to it by a GB. For a GB diffusivity of $10^5 D_{GB,l}$, all of the bubbles near the free surface collapse, which allows additional grain growth near the surface, reducing the number of GBs touching the free surface. As can be seen in Figs~\ref{fig:Impact_GB_Gas_Release_Rate_Over_Time} and \ref{fig:Impact_GB_Number_of_grains}, the drop in the number of grains to 14 that occurs around 10 days coincides with a decrease in the fission gas release rate. A similar slowing of the gas release occurs when the number of GBs touching the surface goes from four to three, at around 40 days. For a GB diffusivity of $10^6 D_{GB,l}$, rapid bubble collapse results in even more rapid grain growth that decreases the number of GBs touching the free surface, resulting in a slowing of the fission gas release. After 60 days, the last intergranular bubble disappears and the fission gas release reduces nearly to zero. 

\begin{figure}[bthp]
\centering
\begin{subfigure}{\textwidth}
  \centering
  \includegraphics[width=0.99\linewidth]{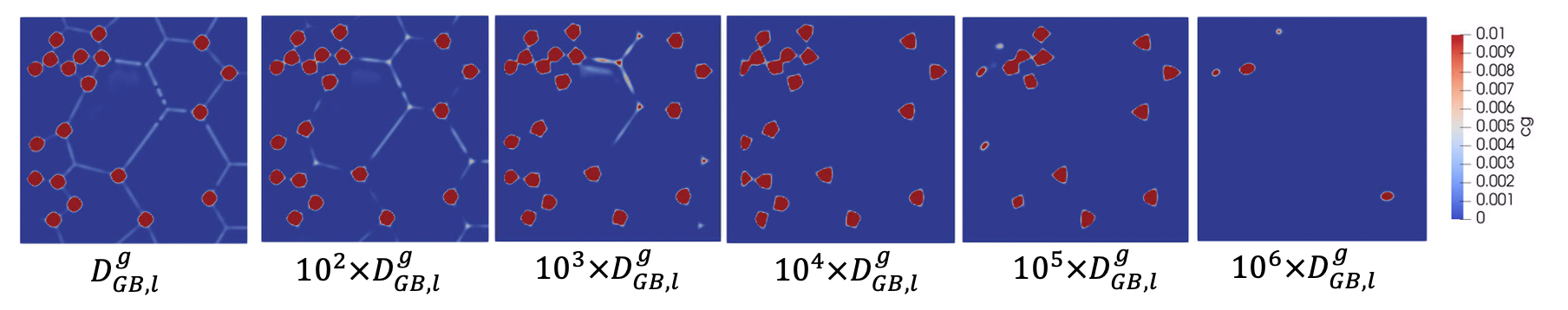}
  \label{fig:Parametric GB Microstructure gas concentration}
\end{subfigure}%
\caption{The Xe concentration in MARMOT across the domain after 100 days for the various GB diffusivities.}
\label{fig:Impact-of-GB-diffusivity-microstructure-gas-concentration}
\end{figure}

From the results of our parametric study, it is possible to draw some conclusions with regards to the correct Xe GB diffusivity. An analysis carried out by Olander and Van Uffelen \cite{Olander_GB_paper} found that a Xe atom would be trapped after a migration distance along a GB equal to the grain size or less. In \cref{fig:Impact-of-GB-diffusivity-microstructure-gas-concentration} we show the gas concentration in MARMOT at the end of our simulations carried out with different GB diffusivities. In our results, the gas depletion region is further than one grain size away from the free surface for GB diffusivities $\geq 10^5 D^g_{GB,l}$. Based on this analysis, we propose that the GB diffusivity should be $\leq 10^4 D^g_{GB,l}$.
%

\subsection{Impact of Surface Diffusivity on Fission Gas Release}
\label{sec:Impact-of-Surface-diffusivity}

To identify how surface diffusion impacts fission gas release, we carry out a parametric study in which we ignore fast GB diffusion and use increasing values of surface diffusivity. As shown in \cref{fig:U-vacancy diffusivity at surface}, at $T = 1600$ K the literature-reported U vacancy surface diffusivity values range from $10^{-12}$ m$^2$/s to $10^{-3}$ m$^2$/s, showing a variation of up to nine orders of magnitude. The value from Zhou and Olander \cite{Zhou_1984} is the highest; therefore, our parametric study begins with the Zhou and Olander surface diffusivity (labeled $D^v_{surf,u}$) and decreases by six orders of magnitude in increments of one order of magnitude. 

\begin{figure}[bthp]
\centering
\begin{subfigure}{.49\textwidth}
  \centering
  \includegraphics[width=0.99\linewidth]{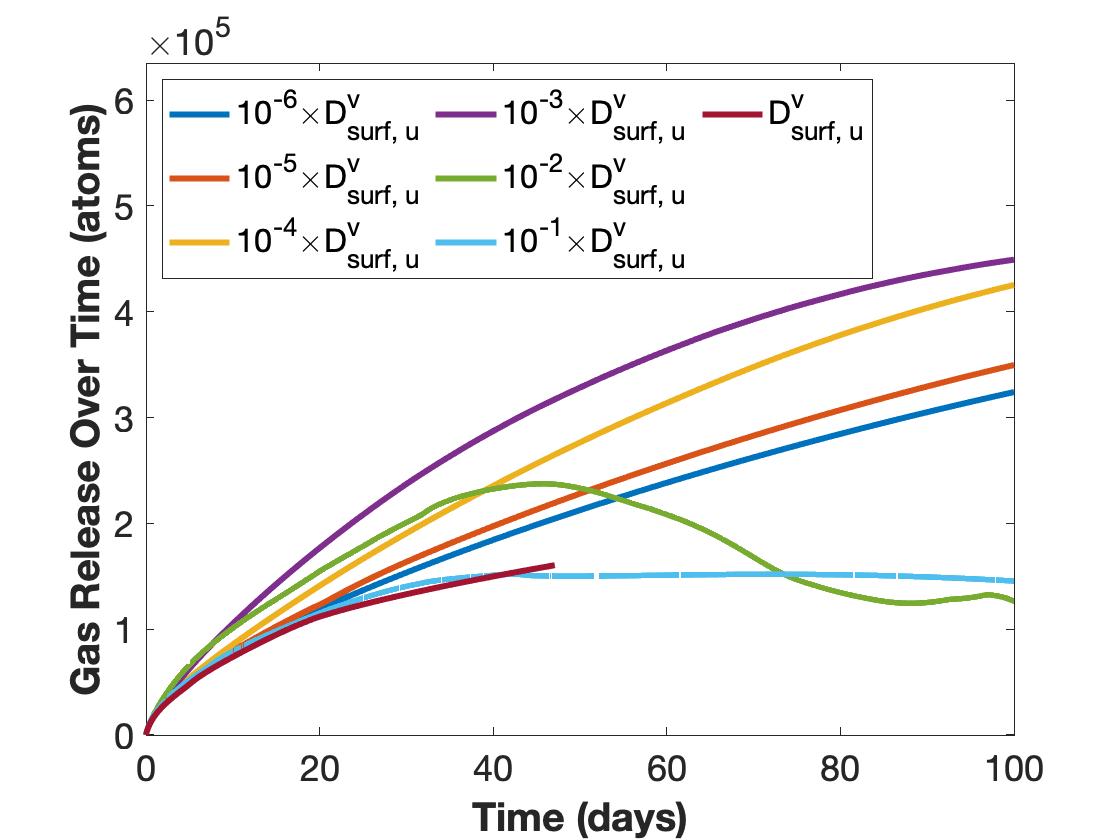}
  \caption{}
  \label{fig:Gas Release Rate Over Time for Parametric Surface}
\end{subfigure}%
\begin{subfigure}{.49\textwidth}
  \vspace{0.5cm}
  \centering
  \includegraphics[width=0.99\linewidth]{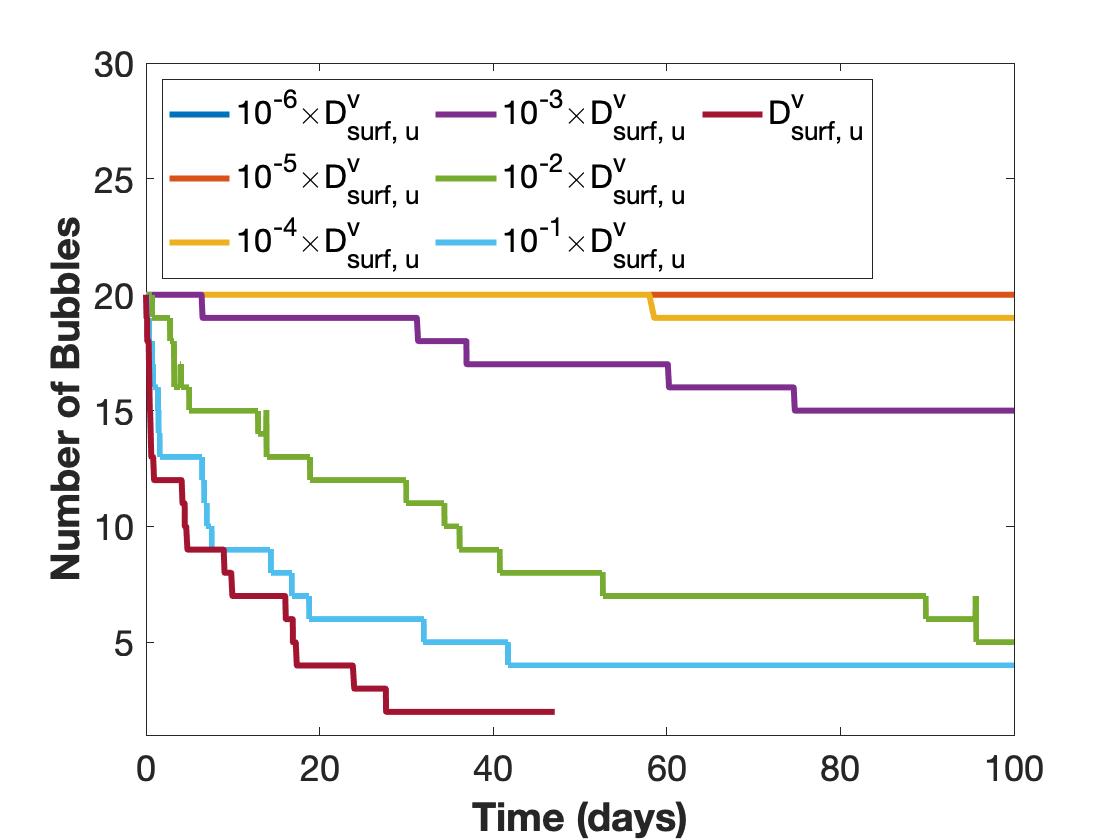}
    \caption{}
\label{fig:Impact_Surface_Number_of_bubbles}
\end{subfigure}%

\begin{subfigure}{.49\textwidth}
  \vspace{0.5cm}
  \centering
  \includegraphics[width=0.99\linewidth]{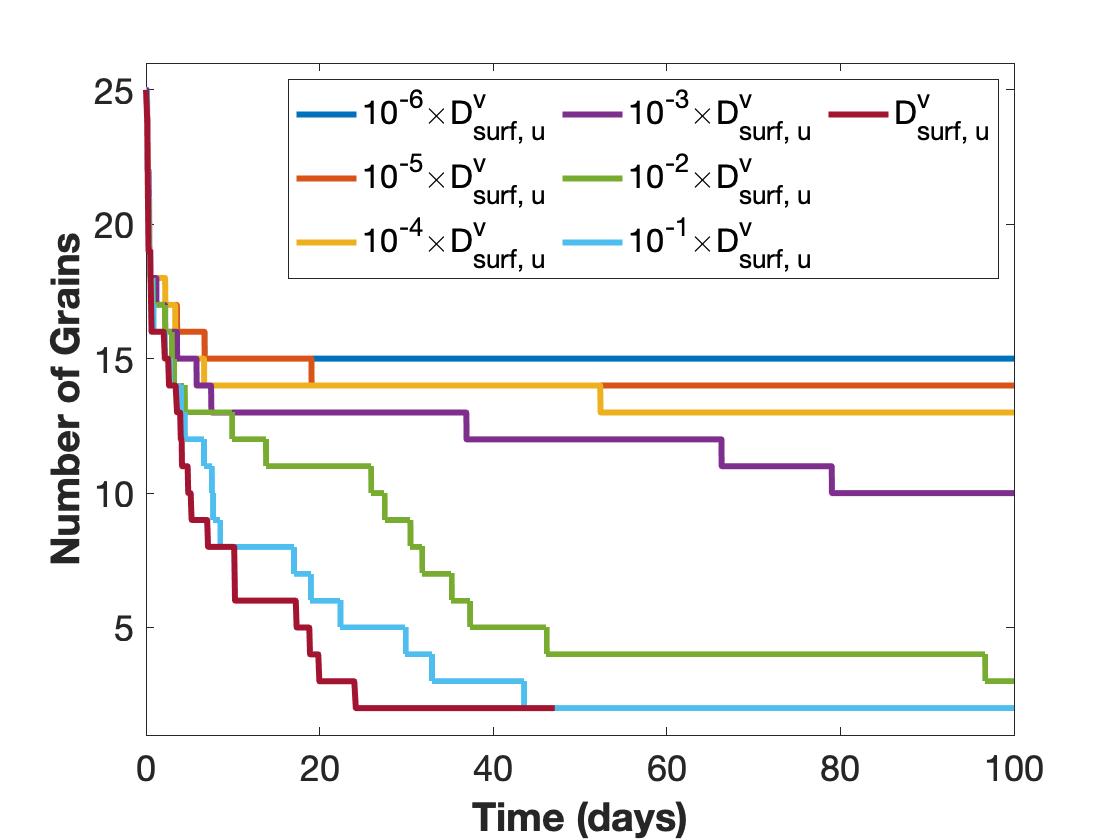}
    \caption{}
  \label{fig:Impact_Surface_Number_of_grains}
\end{subfigure}
\begin{subfigure}{.49\textwidth}
  \vspace{0.5cm}
  \centering
  \includegraphics[width=0.99\linewidth]{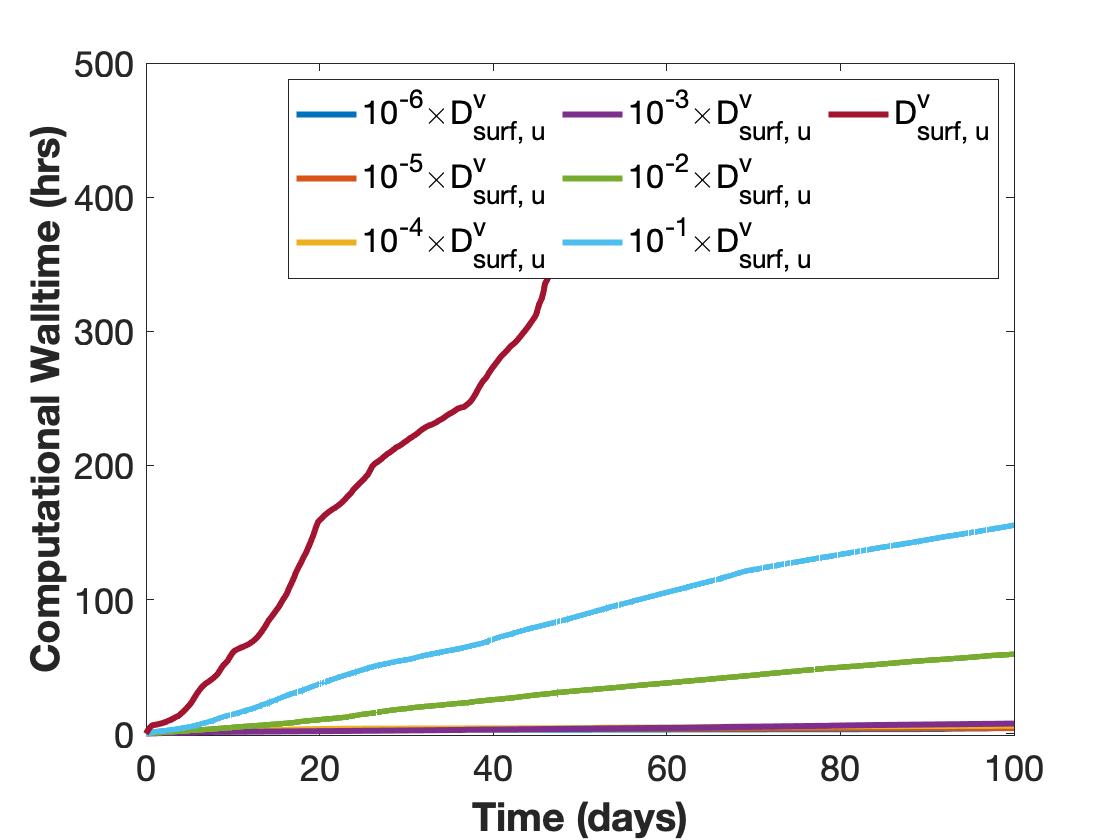}
    \caption{}
  \label{fig:Impact_Surface_Computational_wall_time}
\end{subfigure}
\caption{The impact of surface diffusivity on fission gas release, where (a) shows the gas release rate over time, (b) the quantity of bubbles, (c) the number of grains, and (d) the computational wall time.}
\label{fig:Impact-of-surface-diffusivity}
\end{figure}

Though Section \ref{sec:Impact-of-GB-diffusivity} revealed a large impact of GB diffusivity on fission gas release, this is not the case for surface diffusivity, as shown in \cref{fig:Gas Release Rate Over Time for Parametric Surface}. The magnitude of the gas release for all surface diffusivities is much smaller than for the higher GB diffusivity values. The release increases with increasing surface diffusivity up to a value of $10^{-3}D^v_{surf,u}$. For a diffusivity of $10^{-2}D^v_{surf,u}$, the gas release decreases after around 40 days when depleted regions around bubbles near the free surface result in the nonphysical behavior of gas reentering the fuel. For diffusivities of $10^{-1}D^v_{surf,u}$ and $D^v_{surf,u}$, rapid bubble coalescence and grain growth eliminates the GB paths for diffusion from bubbles to the free surface. This is because there is a large effect of surface diffusivity on number of bubbles (\cref{fig:Impact_Surface_Number_of_bubbles}) and number of grains (\cref{fig:Impact_Surface_Number_of_grains}). An increase in the surface diffusivity results in an increase in bubble mobility, which leads to greater bubble coalescence with neighboring bubbles and less pinning of GBs by bubbles. The increase in the computational wall time is even more dramatic for the high surface diffusivities than for the GB diffusivities, with the total wall time to simulate 50 days with a surface diffusivity of $D^v_{surf,u}$ being 300 times longer than the wall time to simulate 100 days with a surface diffusivity of $10^{-6} D^v_{surf,u}$. 

\begin{figure}[tbhp]
\centering
\begin{subfigure}{\textwidth}
  \centering
  \includegraphics[width=0.99\linewidth]{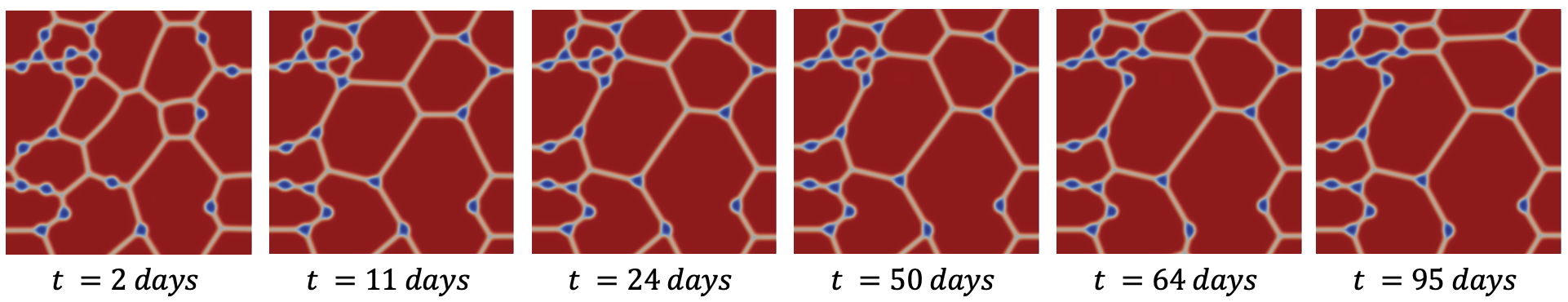}
  \caption{$D_S = 10^{-4} \times D^v_{surf,u}$}
  \label{fig:Parametric surface Microstructure DS=1e-4*D_Zhou}
\end{subfigure}%

\begin{subfigure}{\textwidth}
  \vspace{0.2cm}
  \centering
  \includegraphics[width=0.99\linewidth]{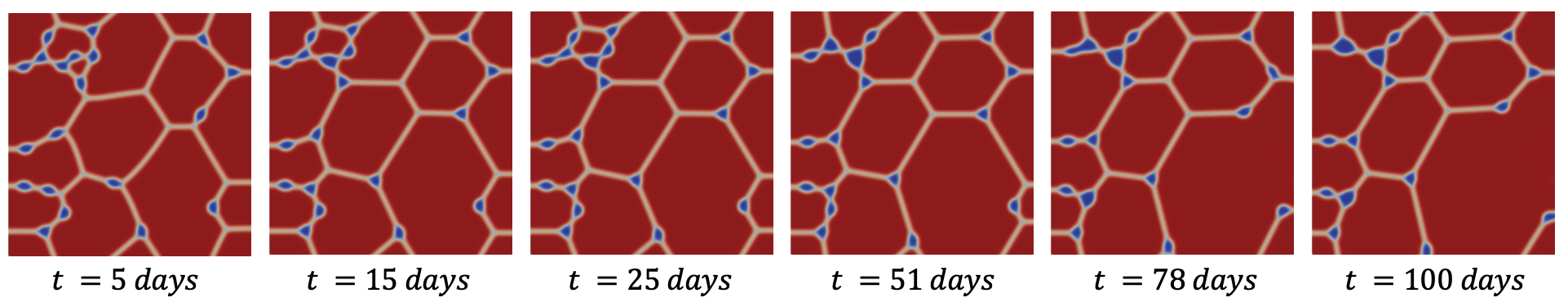}
  \caption{$D_S = 10^{-3} \times D^v_{surf,u}$}
  \label{fig:Parametric surface Microstructure DS=1e-3*D_Zhou}
\end{subfigure}%

\begin{subfigure}{\textwidth}
  \vspace{0.2cm}
  \centering
  \includegraphics[width=0.99\linewidth]{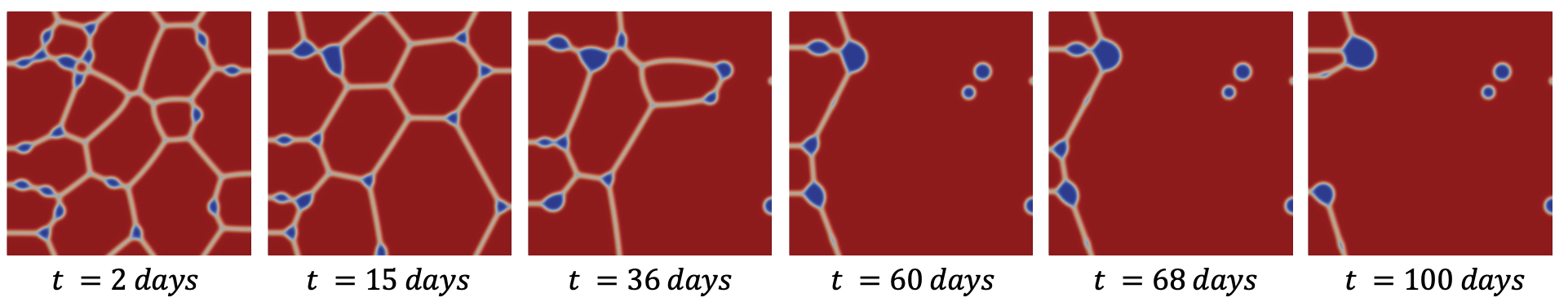}
  \caption{$D_S = 10^{-2} \times D^v_{surf,u}$}
  \label{fig:Parametric surface Microstructure DS=1e-2*D_Zhou}
\end{subfigure}

\begin{subfigure}{\textwidth}
  \vspace{0.2cm}
  \centering
  \includegraphics[width=0.99\linewidth]{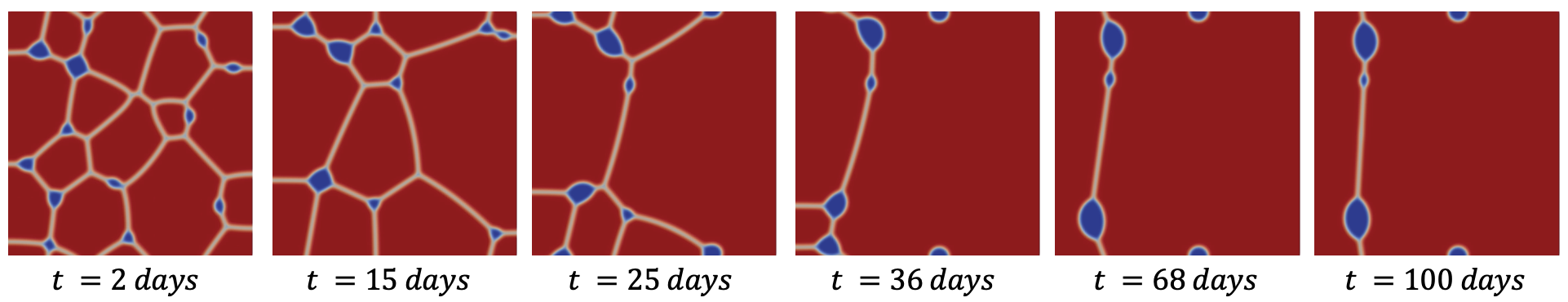}
  \caption{$D_S = 10^{-1} \times D^v_{surf,u}$}
  \label{fig:Parametric surface Microstructure DS=1e-1*D_Zhou}
\end{subfigure}

\begin{subfigure}{\textwidth}
  \vspace{0.2cm}
  \centering
  \includegraphics[width=0.99\linewidth]{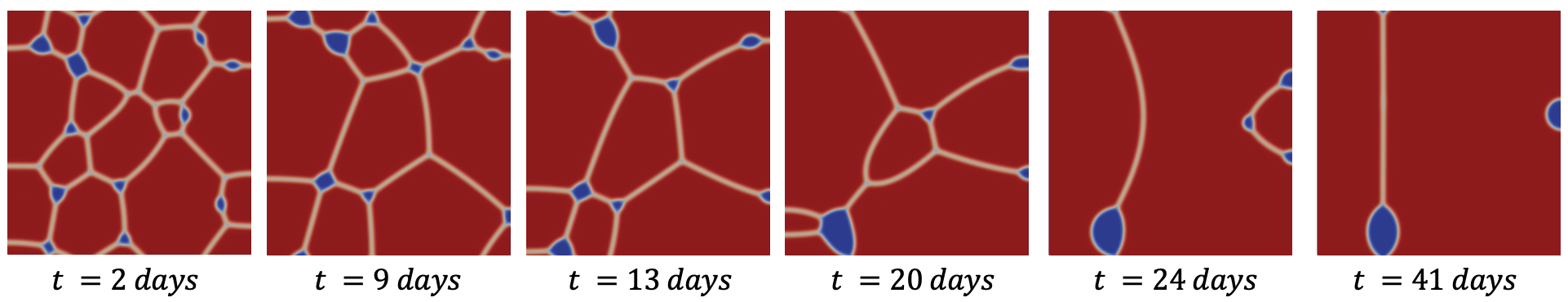}
  \caption{$D_S = D^v_{surf,u}$}
  \label{fig:Parametric surface Microstructure DS=D_Zhou}
\end{subfigure}
\caption{The impact of surface diffusivity on microstructure evolution. Snapshots of the microstructure are shown at different times for the entire range of surface diffusivity values: (a) $D_S = 10^{-4} D^v_{surf,u}$, (b) $D_S = 10^{-3} D^v_{surf,u}$, (c) $D_S = 10^{-2} D^v_{surf,u}$, (d) $D_S = 10^{-1} D^v_{surf,u}$, and (e) $D_S = D^v_{surf,u}$. The times are selected for each snapshot to best illustrate the microstructure change with the different surface diffusivities. The initial microstructure of the simulation is shown in Fig.~\ref{fig:Domain_microstructure}. Grains are shown in red, GBs in yellow, and bubbles in Blue.}
\label{fig:Impact-of-surface-diffusivity-microstructure}
\end{figure}

A qualitative analysis of the microstructure evolution is even more important to understand the impact of surface diffusivity than for the GB diffusivity. The evolution in the microstructure for different surface diffusivities is shown in \cref{fig:Impact-of-surface-diffusivity-microstructure}. Initial grain growth occurs even with lower values of surface diffusivity, as it did for the smaller GB diffusivity cases. No additional microstructure evolution occurred for surface diffusivity $\leq 10^{-5} D^v_{surf,u}$. With a surface diffusivity of $10^{-4} D^v_{surf,u}$, the number of bubbles goes down by one due to bubble coalescence rather than bubble collapse. For surface diffusivities $\geq 10^{-3} D^v_{surf,u}$, bubbles are dragged together by GBs, resulting in additional coalescence. The coalescence further reduces GB pinning and enables additional grain growth. At the end of the simulation, only three grains remain for $10^{-2} D^v_{surf,u}$ and only two remain for $10^{-1} D^v_{surf,u}$ and $ D^v_{surf,u}$ and the remaining bubbles get quite large. 

The amount of bubble coalescence and grain growth that occurs when the surface diffusivity $\geq 10^{-3} D^v_{surf,u}$ seems inconsistent with the percolated intergranular bubbles structures that have been observed in irradiated nuclear fuel (e.g.\ in \cite{White_2004}). However, the interconnected bubble structures that have been shown to form on grain faces are an inherently 3D structure that we cannot observe in our 2D simulations. Therefore, we have performed an additional study on the impact of surface diffusivity on GB and bubble structures using 3D simulations. The computational cost of 3D simulations can be quite high, so we use a reduced domain size of 9 $\mu$m by 9 $\mu$m by 9 $\mu$m with 5 initial grains and 320 gas bubbles with initial radius 384 nm, randomly distributed on the grain boundary faces. The initial mean grain size is approximately 5 $\mu m$. We also only model 75 minutes at $T=1600$ K with a fission rate $\dot{F} = 1.09 \times 10^{19}$ fissions/(m$^3$s). For the 3D analysis, we maintain the same physical and model parameters as in the 2D analysis, except for the absence of the free surface. As our 2D analysis shows that surface diffusivity does not impact fission gas release, our 3D analysis focuses solely on gas-bubble evolution. We carry out two 3D simulations, each with a GB diffusivity of $D_{GB,l}$ and one with a surface diffusivity of $10^{-4} D^v_{surf,u}$ and one of $D^v_{surf,u}$.

\begin{figure}[tbp]
\centering
\begin{subfigure}{.99\textwidth}
  \centering
  \includegraphics[width=0.32\linewidth]{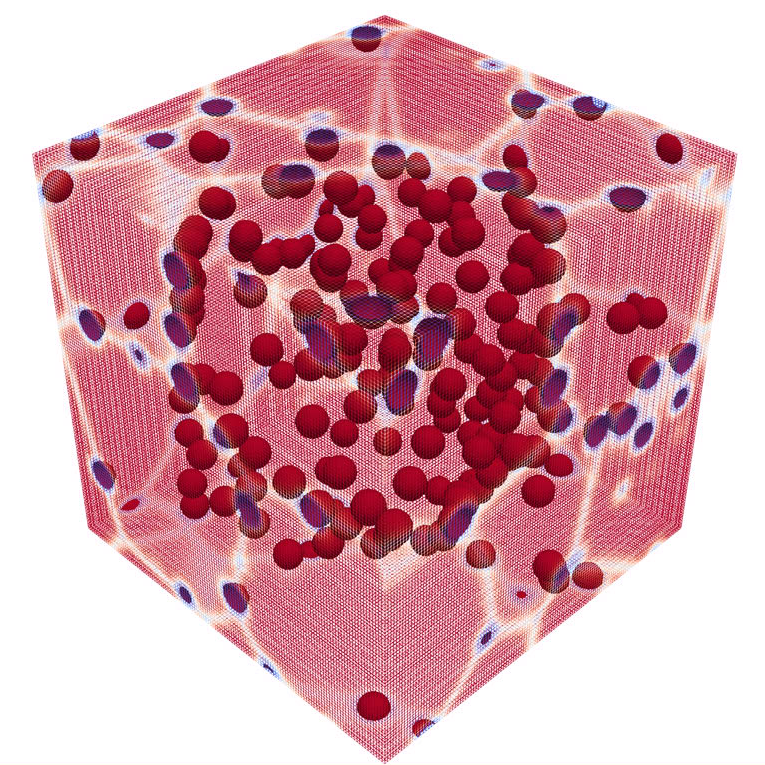}
  \includegraphics[width=0.32\linewidth]{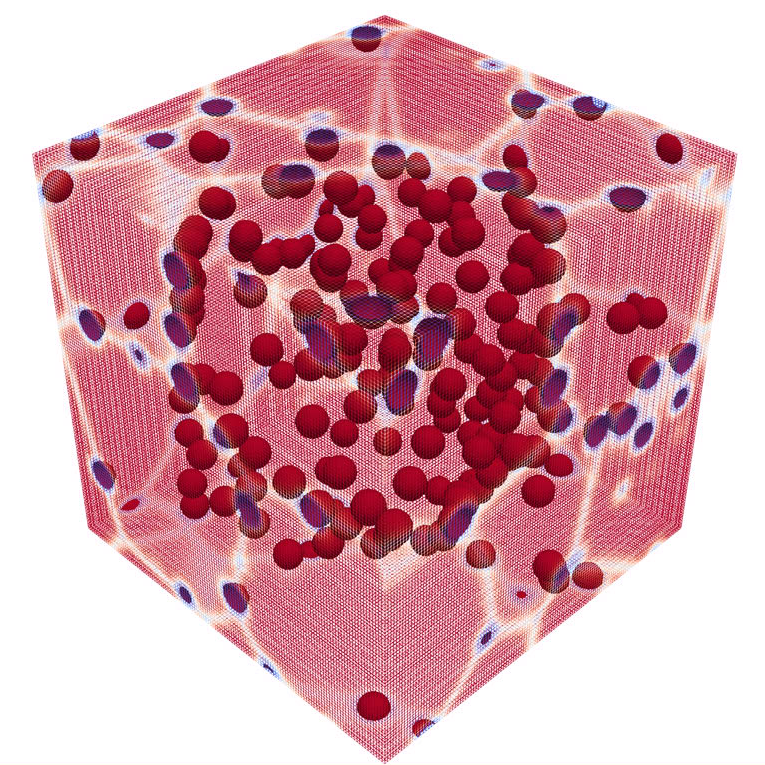}
  \includegraphics[width=0.32\linewidth]{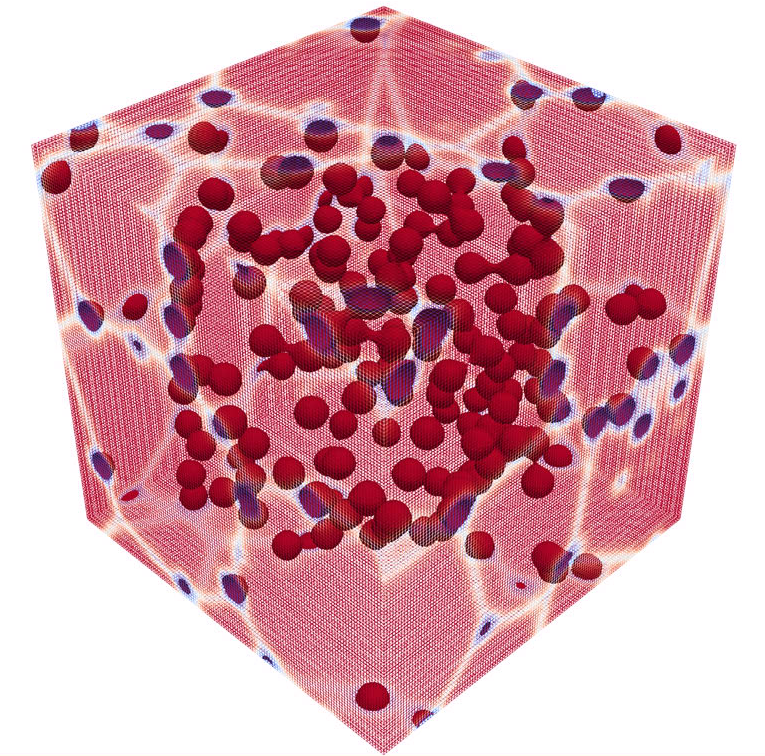}
  \makebox[0.32\linewidth]{$t=1$ min}
  \makebox[0.32\linewidth]{$t=45$ min}
  \makebox[0.32\linewidth]{$t=75$ min}
  \caption{}
\end{subfigure}%

\begin{subfigure}{.99\textwidth}
  \includegraphics[width=0.32\linewidth]{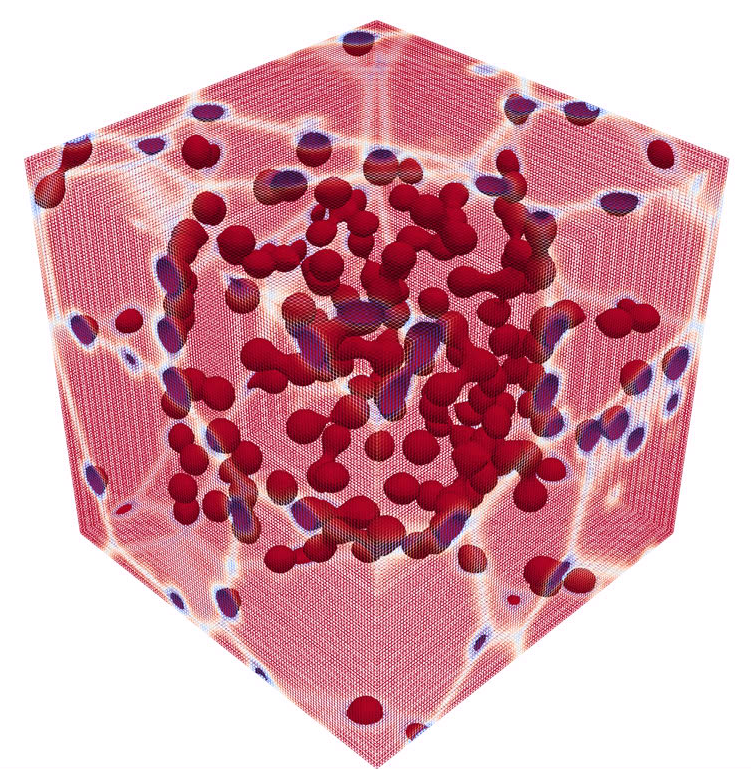}
  \includegraphics[width=0.33\linewidth]{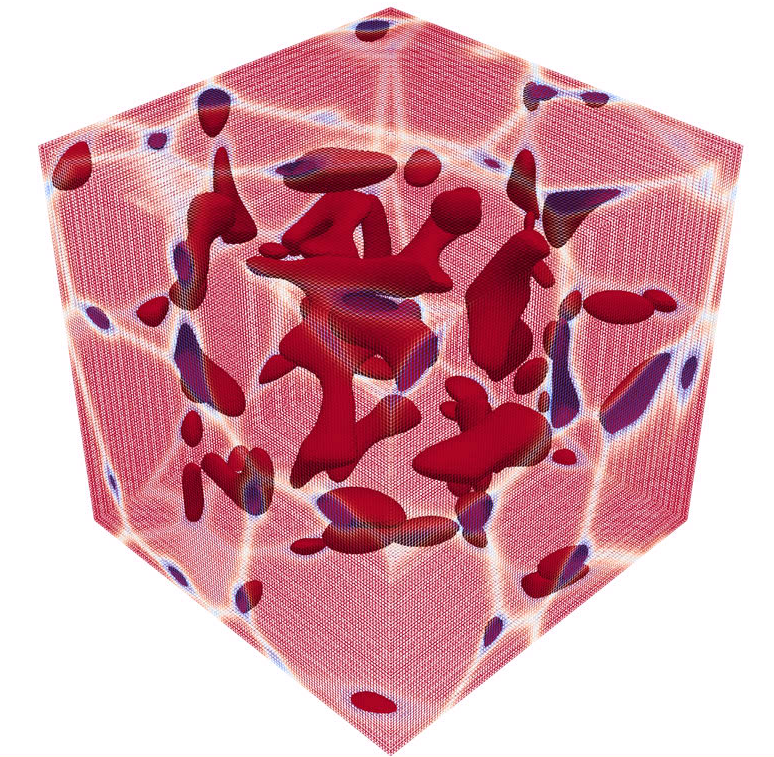}
  \includegraphics[width=0.33\linewidth]{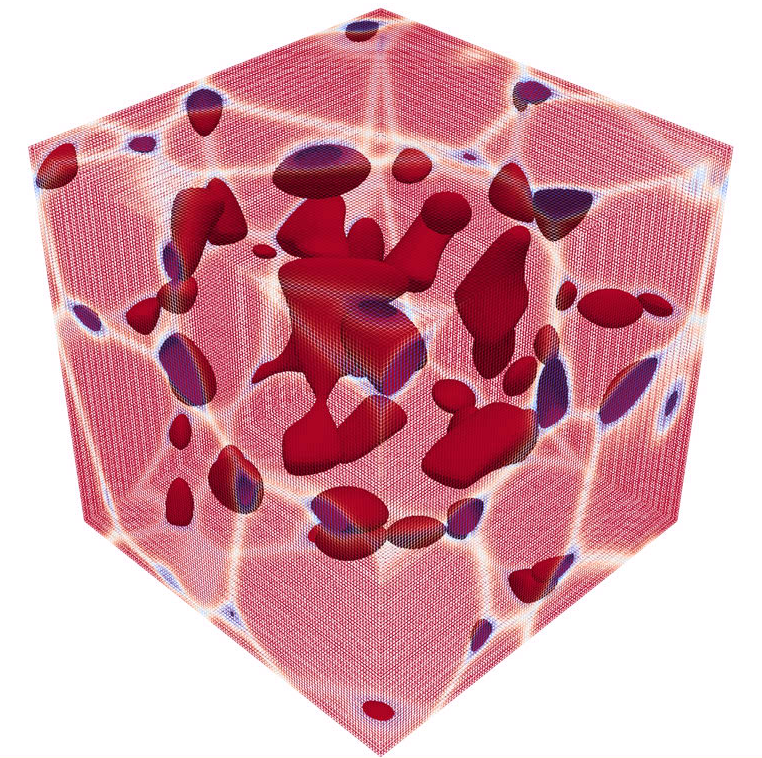}\\
  \makebox[0.32\linewidth]{$t=1$ min}
  \makebox[0.32\linewidth]{$t=15$ min}
  \makebox[0.32\linewidth]{$t=30$ min}\\
  \includegraphics[width=0.32\linewidth]{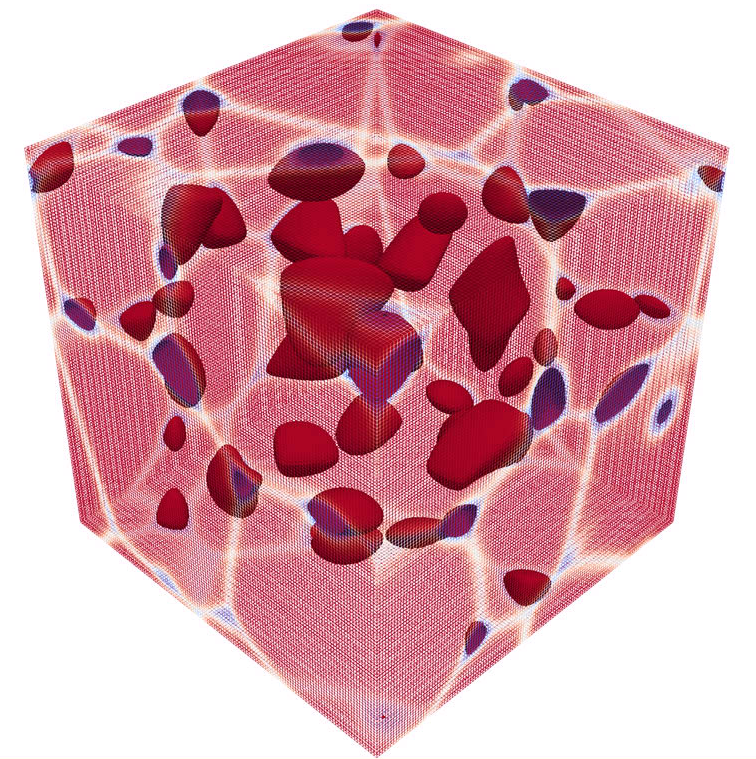}
  \includegraphics[width=0.33\linewidth]{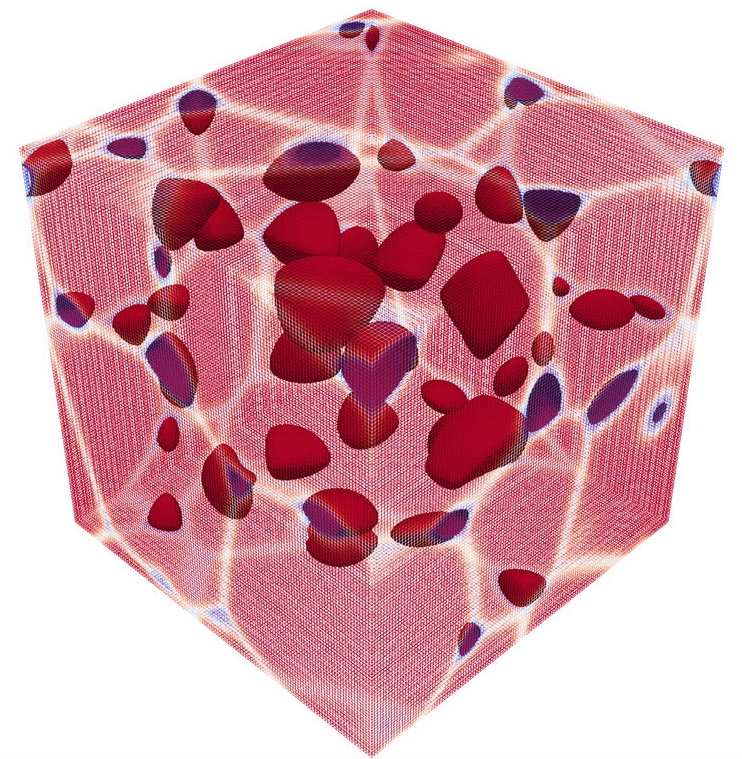}
  \includegraphics[width=0.33\linewidth]{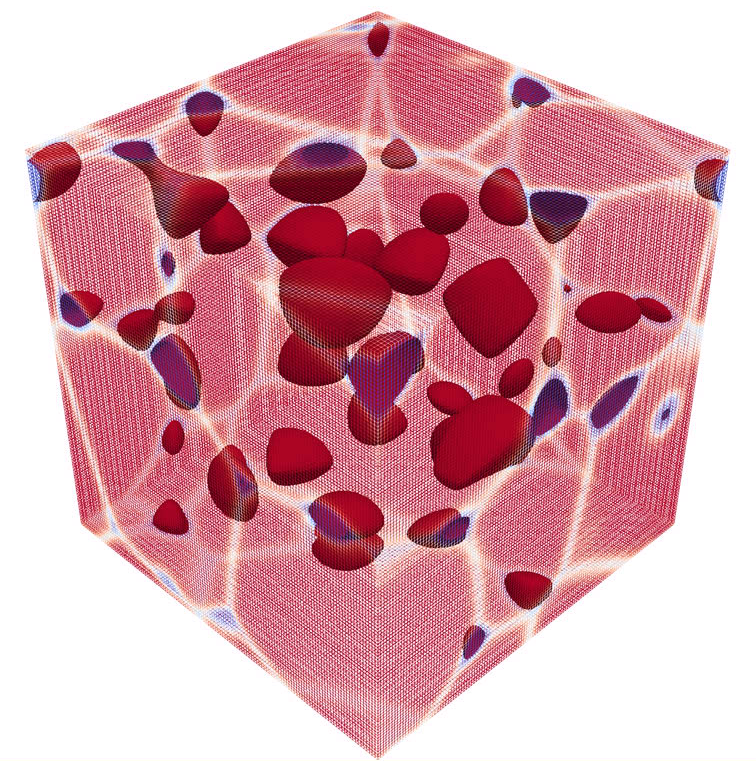}\\
  \makebox[0.32\linewidth]{$t=45$ min}
  \makebox[0.32\linewidth]{$t=60$ min}
  \makebox[0.32\linewidth]{$t=75$ min}
  \caption{}
\end{subfigure}
\caption{Impact of surface diffusion on microstructure evolution in a 9 $\mu$m by 9 $\mu$m by 9 $\mu$m 3D domain. The evolution with a surface diffusivity of $10^{-4} D^v_{surf,u}$ is shown in (a) and with $D^v_{surf,u}$ in (b).}
\label{fig:3D_microstructure_evolution}
\end{figure}

The microstructure evolution in the 3D simulations is shown in \cref{fig:3D_microstructure_evolution}. 
With a surface diffusivity of $10^{-4}D^v_{surf,u}$, most bubbles remain at grain faces, with very few connecting to grain edges. In contrast, with a surface diffusivity of $D^v_{surf,u}$, rapid bubble coalescence occurs. Bubbles initially located at grain faces eventually connect to grain edges or triple junctions due to enhanced surface diffusivity. This movement results in the formation of isolated triple junction bubbles after just 45 minutes. This rate of bubble coalescence is much faster than what is observed in experiments. White \cite{White_2004} suggests that elongated bubbles would not be present in UO$_2$ if morphological relaxation took place under surface diffusion-limited conditions. Our 3D results are consistent with White's suggestion. Based on the results from our 2D and 3D simulations, we suggest that a reasonable bubble surface diffusivity for U vacancies is $\leq 10^{-4}D^v_{surf,u}$.

\section{Discussion}

Based on our 2D simulation results and on the finding from Olander and Van Uffelen \cite{Olander_GB_paper} that gas will only diffuse along GBs for about one grain size before trapping, we suggest that a reasonable value for the Xe GB diffusivity would range from the lower value from Olander and Van Uffelen and up to four orders of magnitude larger. This covers the full range of the two values presented by Olander and Van Uffelen, and is below the GB diffusivity values calculated using MD \cite{Liu_2023}. Thus, we suggest there may be some aspect of the GB structures used in the MD simulations that is resulting in Xe diffusion that is too fast. 

Our results indicate that the GB diffusivity has a large impact on fission gas release, as shown in \cref{fig:Impact_GB_Gas_Release_Rate_Over_Time}. However, it is important to note that our 2D simulations only consider one mechanism for fission gas release: the transport of gas along GBs to the free surface. While this mechanism is present in actual fuel, it is likely not the dominant mechanism for fission gas release. It is commonly assumed that fission gas release primarily occurs due to the three stages discussed in the introduction. However, the interconnection of grain face bubbles and the formation of grain edge tunnels cannot be captured in our 2D simulations. GB diffusivity will have a much smaller impact on these mechanisms. 

Our 2D simulations indicate that surface diffusivity does not have a large impact on fission gas release. However, as mentioned above, our simulations only consider gas release via gas transport along GBs. Surface diffusion will have a large impact on fission gas release via interconnected grain face and grain edge porosity. Large surface diffusivity would result in rapid bubble coalescence that would prevent the formation of interconnected bubble networks, as shown by our 3D simulations in \cref{fig:3D_microstructure_evolution}.

Based on our 2D and 3D simulation results, we suggest that a reasonable bubble surface diffusivity for U vacancies is four orders of magnitude below the value from Zhou and Olander \cite{Zhou_1984} or less. Thus, we recommend surface diffusivities similar to those obtained using mass transport methods \cite{Marlowe_1968,Henney_1968,maiya1971surface,Amato_1966} or lower. However, more information regarding surface diffusivity of fission gas bubbles is needed.

White and Tucker \cite{White_2004} found that assuming surface diffusion-limited conditions \cite{White_2004} would result in rapid bubble coalescence, in contrast to the elongated bubble structures observed in their fuel characterization. Our results are also consistent with this finding. This suggests that bubble surfaces may behave differently than the free surfaces used to measure surface diffusion. This is consistent with discrepancies between the measured surface to GB energy ratios (for free surfaces) and the bubble semi-dihedral angles that are observed in irradiated fuel. The surface to GB energy ratio is typically found to be around two  \cite{hodkin1980ratio}, yet semi-dihedral angles in fuel indicate energy ratios ranging from 0.6 to 0.8. 

Since gas bubble surfaces exhibit distinct behavior, it is crucial to investigate the differences between UO$_2$ free surfaces and gas bubble surfaces. One potential approach would be to conduct MD simulation studies focused on bubble surfaces to understand surface diffusion around these areas. Existing experimental studies measuring surface diffusivity in UO$_2$ focus on free surfaces rather than gas bubble surfaces. Approaches that focus on the gas bubble surfaces or bubble mobility are needed. For example, an in-situ Xe-implantation experiment in UO$_2$ could be carried out to observe bubble evolution and measure the kinetics of bubble coalescence, providing insight into gas bubble surface diffusivity.

Finally, the 2D simulations carried out in this work have provided initial insights into the importance of considering GB and surface diffusion in phase field models of fission gas bubble evolution. This behavior has not been considered in past phase field models of fission gas release \cite{DongUk-Kim, Larry_2019, Millet_2011, Millet_2012_a, Millet_2012_b, Millet_2012_c, Yulan-Li_2013, Xiao_2020}, but here we have shown the large impact it has on the transport of gas atoms along GBs and on the bubble mobility, which impacts both bubble coalescence and bubble pinning of GBs. Future 3D simulations are needed that consider GB and surface diffusion to investigate grain face bubble interconnection and grain edge tunnel formation. 

\section{Conclusion}
In this study, we utilized the hybrid phase field/cluster dynamics model developed by Kim et al.~\cite{DongUk-Kim} to  investigate the impact of fast GB and surface diffusion on fission gas bubble and GB evolution, as well as fission gas release, which has been typically neglected in mesoscale fission gas models. Our review of bulk, GB, and surface diffusivity values for U vacancies and Xe in UO$_2$ showed significant uncertainty due to the large range of values. Thus, we carried out parametric studies on the impact of GB and surface diffusivity on fission gas behavior using a modified version of the hybrid model. Our parametric study on the impact of GB diffusivity found that it had a large impact on the fission gas release in our 2D domain. The results indicated that the GB diffusivity is likely $\leq 10^4 D^g_{GB,l}$, where $D^g_{GB,l}$ is the lower grain boundary diffusivity value rerported by Olander and Van Uffelen~\cite{Olander_GB_paper}. Our parametric study on the impact of surface diffusivity found that it had a large impact on bubble coalescence and GB migration. Our 2D and 3D results indicate that the surface diffusivity is likely $\leq 10^{-4} D^v_{surf,u}$, where $D^v_{surf,u}$ is the surface diffusivity value reported by Zhou and Olander~\cite{Zhou_1984}. However, additional experiments and atomistic simulation results are needed to reduce the uncertainty on the GB and surface diffusivities in UO$_2$ and 3D simulations are needed to further explore the fission gas behavior.

\section*{CrediT Authorship Contribution Statement}
\textbf{Md Ali Muntaha}: Investigation, Methodology, Analysis, Software,
Writing – original draft. \textbf{Sourav Chatterjee}: Investigation, Writing – review and editing. \textbf{Sophie Blondel}: Conceptualization, Writing – review and editing. \textbf{Larry Aagesen}: Conceptualization, Investigation, Supervision,Writing – review and editing. \textbf{David Andersson}: Conceptualization, Investigation, Supervision, Writing – review and editing. \textbf{Brian D. Wirth}: Conceptualization, Writing – review and editing. \textbf{Michael R. Tonks}: Conceptualization, Funding acquisition, Project administration, Supervision, Writing – review and editing.

\section*{Declaration of Competing Interest}
The authors declare that they have no known competing financial interests or personal relationships that could have appeared to influence the work reported in this paper.

\section*{Acknowledgements}
We express our gratitude for the high-performance computing resources provided by the University of Florida's HiPerGator clusters, which facilitated the execution of computationally intensive 2D simulations. Additionally, we acknowledge the support from leadership-class computers, such as INL Sawtooth \cite{Parker_2023}, for running 3D large scale simulations. 

This material is based upon work supported by the U. S. Department of Energy, Office of Nuclear Energy and Office of Science, Office of Advanced Scientific Computing Research through the Scientific Discovery through Advanced Computing (SciDAC) project on Simulation of Fission Gas through the grant DOE DE-SC0018359 at the University of Tennessee.


\section*{Data Availability}
The MOOSE input files used to generate the simulation results in this paper can be obtained from the authors upon reasonable request.

\bibliographystyle{elsarticle-num.bst}
\bibliography{My_Collection}

\newpage

\end{document}